\documentclass[12pt]{article}
\usepackage[a4paper, portrait, margin=1.1811in]{geometry}
\usepackage[english]{babel}
\usepackage[utf8]{inputenc}
\usepackage[T1]{fontenc}
\usepackage{helvet}
\usepackage{etoolbox}
\usepackage{graphicx}
\usepackage{titlesec}
\usepackage{caption}
\usepackage{booktabs}
\usepackage{xcolor} 
\usepackage[colorlinks, citecolor=black]{hyperref}
\usepackage{caption}
\graphicspath{ {./images/} }
\usepackage{scrextend}
\usepackage{fancyhdr}
\usepackage{graphicx, subfigure}
\usepackage{setspace}
\usepackage{enumitem}
\usepackage{lipsum}
\usepackage{amsmath}
\usepackage{amssymb}
\usepackage{longtable}
\usepackage{float}
\usepackage{tabularx}
\usepackage{xltabular}
\usepackage{mathptmx}%
\usepackage{tabularx}
\usepackage{color}
\usepackage{titlesec}
\usepackage{soul,xcolor}
\usepackage{cancel,xcolor}
\allowdisplaybreaks
\usepackage{makecell}
\usepackage{grffile}
\usepackage{pdflscape}

\newcolumntype{L}[1]{>{\raggedright\let\newline\\\arraybackslash\hspace{0pt}}m{#1}}
\newcolumntype{C}[1]{>{\centering\let\newline\\\arraybackslash\hspace{0pt}}m{#1}}

\newlist{steps}{enumerate}{1}
\setlist[steps, 1]{label = Step \arabic*:}

\usepackage{natbib}

\fancypagestyle{plain}{
	\fancyhf{}

}

\makeatletter
\patchcmd{\@maketitle}{\LARGE \@title}{\fontsize{16}{19.2}\selectfont\@title}{}{}
\makeatother

\usepackage{authblk}

\newsavebox\affbox

\author[1]{\textbf{Sandip K Pal}}
\author[1]{\textbf{Arnab Koley}}
\author[1]{\textbf{Pritam Ranjan}$^*$}
\author[2]{\textbf{Debasis Kundu}}
\affil[1]{OM\&QT Area, Indian Institute of Management Indore, MP, India
}
\affil[2]{Department of Mathematics and Statistics, Indian Institute of Technology Kanpur, UP, India
}

\titleformat{\subsection}[runin]
  {\normalfont\bfseries}{\thesubsection}{1em}{}[:]
  
\title{\textbf{\huge {Modeling time to failure using a temporal sequence of  events}}}

\date{}    
\begin{document}
\setstcolor{red}

\setcounter{page}{1}
\renewcommand{\thepage}{\arabic{page}}
	
\captionsetup[figure]{labelfont={bf},labelformat={default},labelsep=period,name={Figure }}	\captionsetup[table]{labelfont={bf},labelformat={default},labelsep=period,name={Table }}
\maketitle
\noindent\rule{15cm}{0.5pt}
\begin{abstract}
In recent years, the requirement for real-time understanding of machine behavior has become an important objective in industrial sectors to reduce the cost of unscheduled downtime and to maximize production with expected quality. The vast majority of high-end machines are equipped with a number of sensors that can record event logs over time.  In this paper, we consider an injection molding (IM) machine that manufactures plastic bottles for soft drink. We have analyzed the machine log data with a sequence of three type of events, “running with alert”, “running without alert”, and “failure”. Failure event leads to downtime of the machine and necessitates maintenance. The sensors are capable of capturing the corresponding operational conditions of the machine as well as the defined states of events. This paper presents a new model to predict a) time to failure of the IM machine and b) identification of important sensors in the system that may explain the events which in-turn leads to failure. The proposed method is more efficient than the popular competitor and can help reduce the downtime costs by controlling operational parameters in advance to prevent failures from occurring too soon. \\ \\
		\let\thefootnote\relax\footnotetext{
			\small $^{*}$\textbf{Corresponding author:} Pritam Ranjan, \textit{
	\textit{E-mail address: \color{black}pritamr@iimidr.ac.in}}\\	
		}
	\end{abstract}
 
\noindent\rule{15cm}{0.4pt}

\doublespacing

\section{Introduction}\label{introduction}
Running manufacturing equipment involves maintenance of machines on a regular basis. Some machinery like health care devices (CT scanner, MRI, etc.), Xerox machines, computers require a proactive or preventing servicing at regular intervals. Although a popular and well accepted approach, preventive maintenance tasks are carried out according to a timetable, and not always when the equipment specifically calls for them. In order to arrange for an efficient preventive repair activity, it is crucial to predict machine failures with enough lead time.

Industry 4.0 brings forth intelligent machines equipped with sophisticated sensors, embedded software, and robotics which gather and store data as machine logs in a  semi-structured format. These data are usually collected while machine is in running condition, and primarily consists  of operation events, performance counters, and alert messages, among others.  Analyzing the  data collected from sensors on the factory floor provides real-time visibility of the machines and helps in performing predictive maintenance to extend the equipment runtime. 
%
%
The log data can be utilized to comprehend the causal connections between the sequence of system generated alert messages and their operating conditions, as well as to predict the likelihood of a failure event. The example provided by \cite{Ding} illustrates a concrete machine that produces a series of operating events (i.e., warm-up, landing leg unfolding, cantilever unfolding, and concrete pumping) for a concrete pump track during a single run-time. According to \cite{Zhou}, a computerized tomography scan machine can produce a number of events pertaining to different machine behaviors and activities, system failures, operator or user actions, the status of a subsystem task, etc. Another example of sequential data of patient conditions can be captured using electronic health record (EHR) data from wearable devices, hospital records, medical test and diagnosis etc. Thus, an individual's EHR consists of two types of data: static data, which consists of demographic variables that are considered fixed for the majority of analyses, and sequential data, which consists of the sequence of medical codes that describe a person's medical history \citep{Glazier}.

%

Several predictive models have been proposed by different authors to predict the failure time using the sequence of events. \cite{Zhou} developed a Cox-proportional hazard (CPH) model to predict the time to failure model, which is built on a set of previously happened events called triggered events that may result in machine failure. This model can be applied in real-time to predict system failure based on preemptive event triggers. A comparison between temporal event incidents and time series data of services, operations, etc. obtained from telemetry system have been analyzed in \cite{Luo} to identify the root cause of an incident in a system. \cite{Luo} proposed a framework which consists of three stages: data pre-processing, event extraction, and correlation analysis. In the data pre-processing stage, the authors use a sliding window to transform raw time series data into a matrix of features. In the event extraction stage, they use a combination of predefined rules and machine learning techniques to identify events that occurred in the system. During the correlation analysis stage, a correlation-based algorithm is employed to determine the occurrences that exhibit the highest level of association with the incident. A few other works on the correlation based event prediction model are \cite{Motahari,Philip,Zhu,Jian-Guang}. \cite{Agrawal} use association mining to learn a pattern based on a historical sequence of past events to predict the probable occurrence of next event(s). In retail sector, the market basket analysis has been recognized as a proven and successful application of association rule mining for cross selling, product placement, promotion affinity analysis, and product promotion and targeting \citep{Kohavi}, for mining gene sequence expression \citep{Jiang} and for web-log mining \citep{Huang, Rudin}. 



\begin{figure}[!h]
	\centering
	\includegraphics[width=1.0\textwidth]{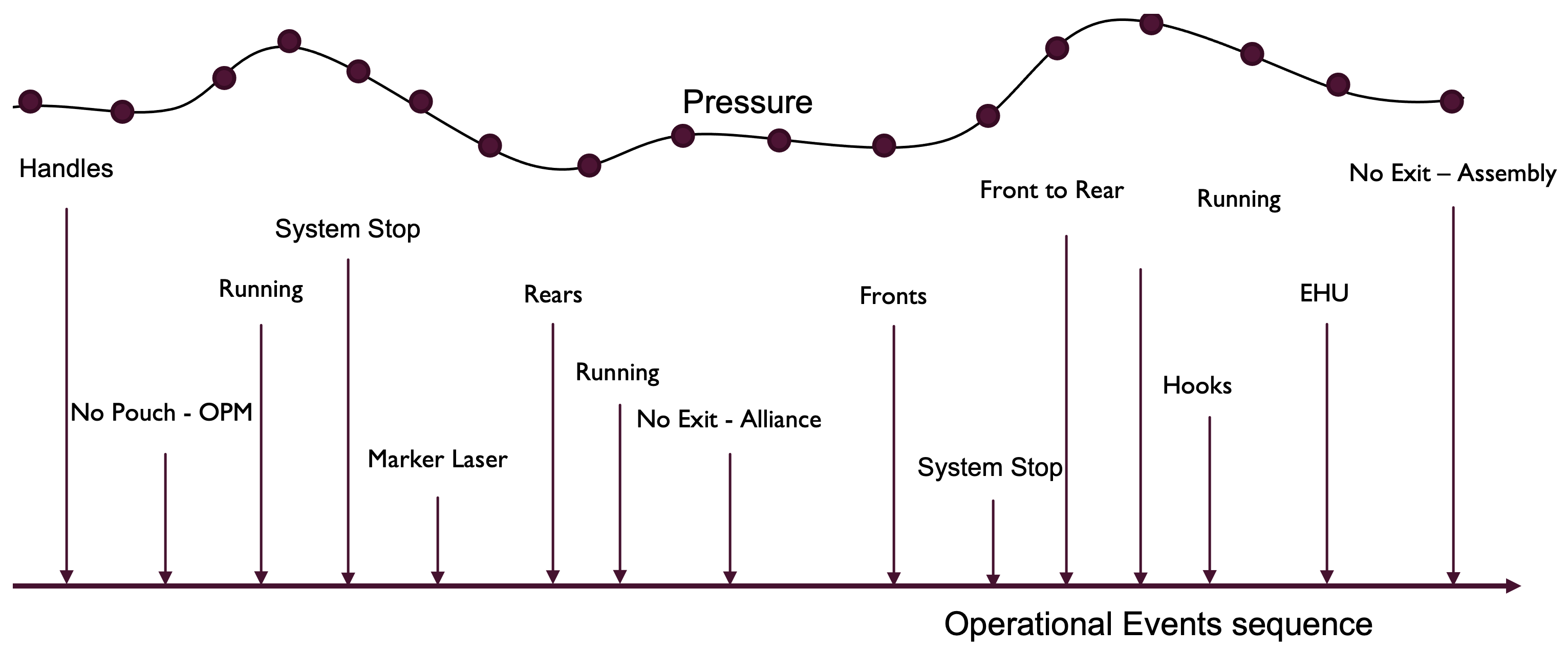}
	\caption{An example of sequence of events while machine is in operational condition.}
	\label{fig1}
\end{figure}

The main objective of this study is to propose a time to failure model of an injection molding (IM) machine for a plastic soft drink bottle. Figure~\ref{fig1} depicts  different operational sequence of a few events of an IM machine which typically provide sufficient information for engineers to diagnose the working condition of equipment.

The machine is said to be in running operative condition until a failure occurs which forces the machine to stop working. In the running condition, one can observe either of the two events, viz. ``running without alert'' and ``running with alert'' to exist at any time point. An analogous example of this kind of state can be compared for a modern car. The car is in motion, but a warning sign, such as ``Seat Belt Warning Sign'' or ``Non-Economical Driving'' or ``Take a Coffee Break'' is being displayed on the dashboard. Note that we have defined an ``epoch'' as the time gap between when the machine starts working after system maintenance till the time of next failure. Note that the time between two consecutive failures is downtime of the system plus an epoch.  

\begin{figure}[H]
	\includegraphics[scale=0.52]{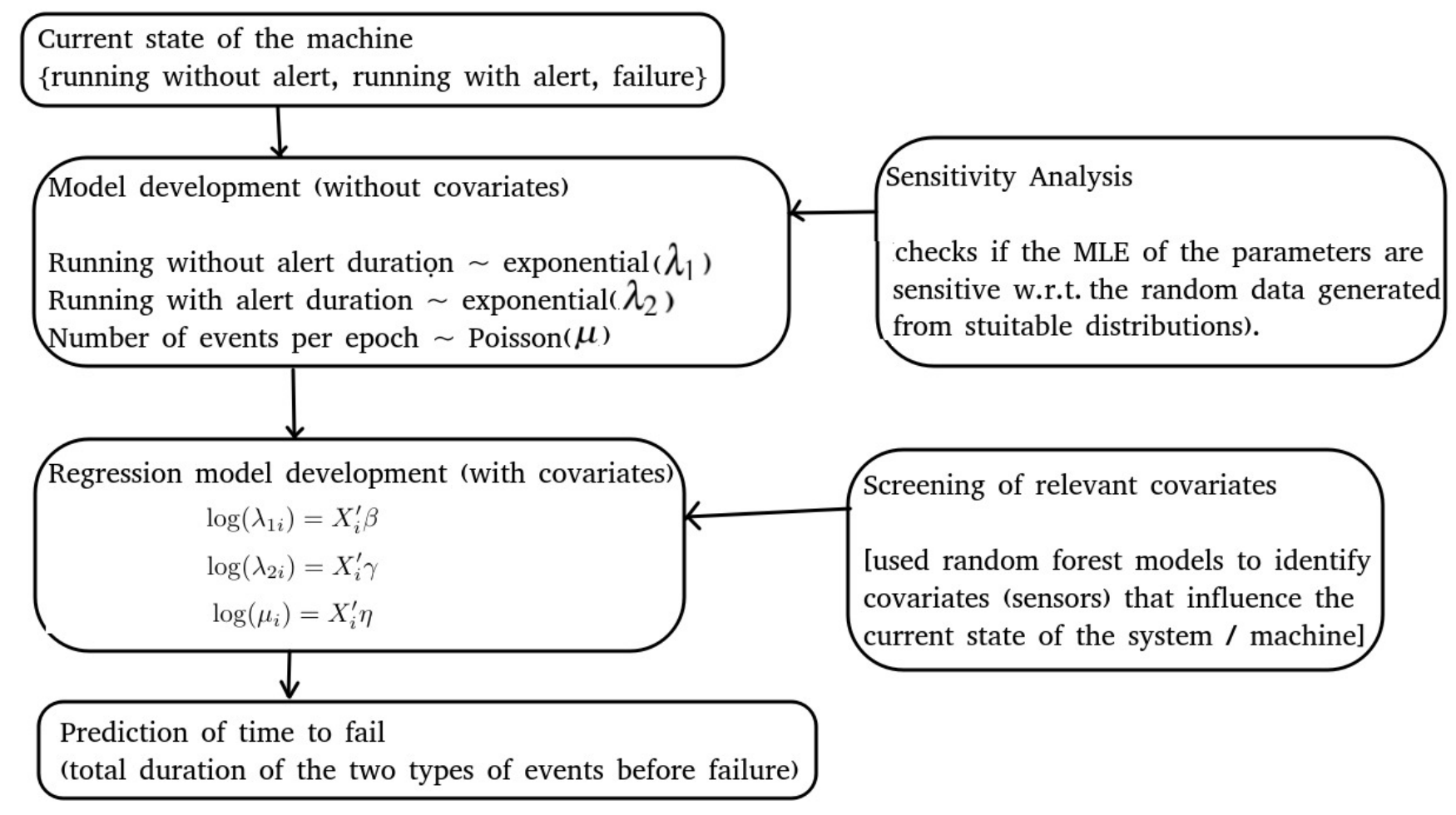}
	\caption{Roadmap of the proposed methodology for predicting time to failure.}
	\label{diagram_roadmap}
\end{figure}

Motivated by the actual need for accurate diagnosis of the working condition of an IM machine, we propose a parametric method to predict the time to failure. The time spent by a machine on a random incidence of ``running without alert'' state is assumed to follow an  exponential distribution with rate parameter $\lambda_1$, whereas the time duration on a random ``running with alert'' event follows an independent exponential distribution with rate parameter $\lambda_2$.  We also assume that the total number of  events in an epoch follows a shifted Poisson distribution. The distributional assumptions are based on popular practitioners' choice and are easy to implement and interpret. The proposed model also turns out to work quite satisfactorily as compared to the popular CPH model to predict failure time of the machine. For our IM machine application, there are 72 sensors, and the key objectives of this study include the identification of key sensors that may have some insights on the reason behind the failure. We use popular machine learning (ML) classifier called the random forest for the identification of important sensor based covariates, which were subsequently used to build a model to predict time to failure. 
{A detailed flowchart of the proposed methodology is presented in Figure~\ref{diagram_roadmap}}.
	
Section~\ref{Machine & Data description} discusses the salient features and summary statistics of the IM machine. Event level model (without the sensor data) has been developed in Section~\ref{model description and estimation}. Next,  Section~\ref{Analysis of data without covariates} discusses simulation study results to assess the sensitivity of the parameters estimates of the proposed model. Section~\ref{Modeling and analysis of sensor data} builds the predictive model using the data on 72 sensors and the event level model in  Section~\ref{model description and estimation}. Accurate prediction of time to failure using the proposed model has been discussed in Section~\ref{Prediction of Time to Failure}. We use a variety of goodness of fit measure to show that the proposed method exhibit higher prediction accuracy as compared to the popular CPH model. Finally, the concluding remarks and possible future directions are discussed in Section~\ref{Conclusion}.

\section{IM machine and data description}\label{Machine & Data description}
\begin{figure}[!h]\centering
		\includegraphics[scale=0.85]{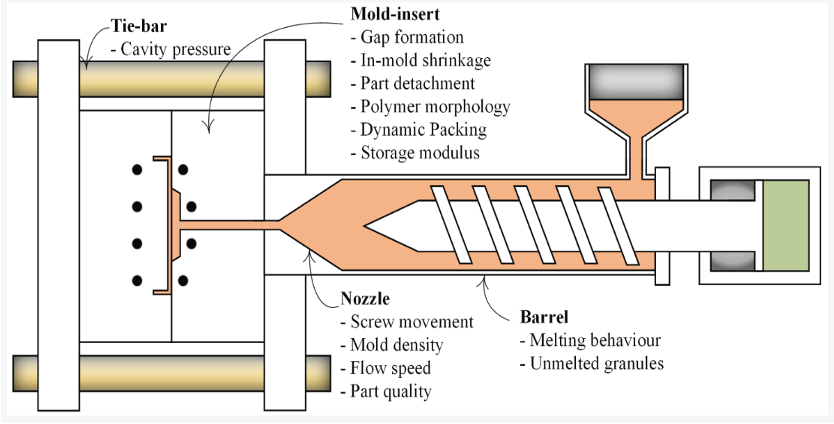}
	\caption{An injection molding (IM) machine schematic diagram (source: https://prototechasia.com/en/injection-molding/questions-injection-molding)}
	\label{Fig:fig2}
\end{figure}

In this section we discuss the IM machine and the data observed from therein. An IM machine is typically  used to manufacture items such as plastic bottles, crates, jars, etc. The IM machine application that motivated this study produces soft drink bottles. In general, an IM machine is made up of two units, the injection unit and the clamping unit (depicted in Figure~\ref{Fig:fig2}). The plastic granules are kept in the hopper which go directly into the barrel. The granules are melted in heat chambers, and the screw injects the melted plastic into the clamping unit of the machine through a nozzle. In the clamping unit, there is a removable mold that can be placed with a mold cavity where the melted plastic will be injected, and the machine will produce bottles. The removable mold can have different shapes to produce different items from this machine. After producing bottles, it goes through an automated quality checking process to reject or accept the item as per the quality control parameters. This research focuses on the system logs data that are captured through various sensors mounted in the IM machine. 

\begin{figure}[h]
	\centering
	\includegraphics[width=1.0\textwidth]{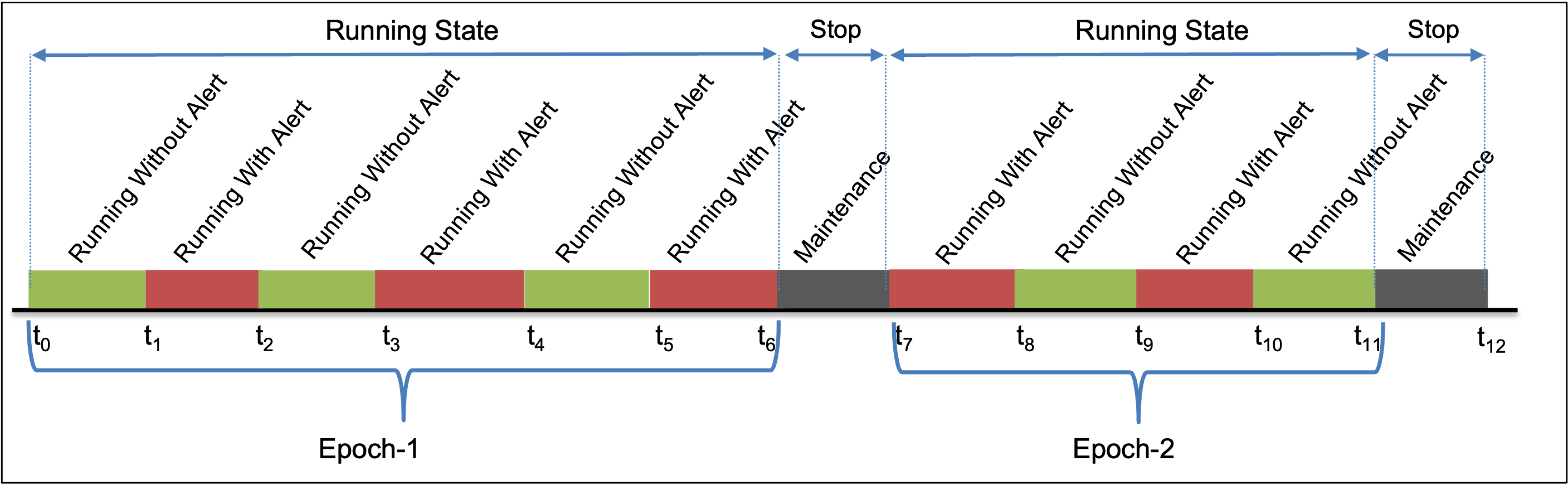}
	\caption{Illustration of a snapshot of different states of the machine (i.e., ``running without alert'', ``running with alert'' and ``failure'').}
	\label{fig4}
\end{figure}

Our data was collected over six days. The various events being experienced by the machine have been divided into three categories: ``running without alert'', ``running with alert'' and ``failure''. It is to be noted that ``running with alert'' refers to a condition of the equipment where it is operational but giving an alert or warning message. The machine's ``failure" condition suggests that it has  stopped functioning and requires maintenance. The graphical representation of the data structure is illustrated in Figure~\ref{fig4}, where the machine starts working at time point $t_0$ with ``running without alert'' event, and then alternates between ``running with alert'' and ``running without alert'' until $t_6$, where it hits ``failure" and stops working. The time duration between $t_0$ and $t_6$ is defined as one \textit{epoch}. The machine undergoes a maintenance work from time point $t_6$ to $t_7$. Subsequently, the machine starts working at time point $t_7$ in ``running with alert'' state till time point $t_8$ at which it starts experiencing ``running without alert'' event. This process continues till $t_{11}$ before another failure occurs, at which point it needs to be serviced once more. The entire time duration from $t_7$ to $t_{11}$ forms another epoch. The entire operational period continues with this process in the same manner. Note that the events ``running without alert'' and ``running with alert'' occur alternatively followed by ``failure" finally. It may also be possible that the machine experiences only one type of events, say ``running without alert'' or ``running with alert'' in an epoch before ``failure". In this illustrative image only two epochs are shown for the purpose of understanding.

\begin{figure}[H]
	\centering
	\includegraphics[width=0.95\textwidth]{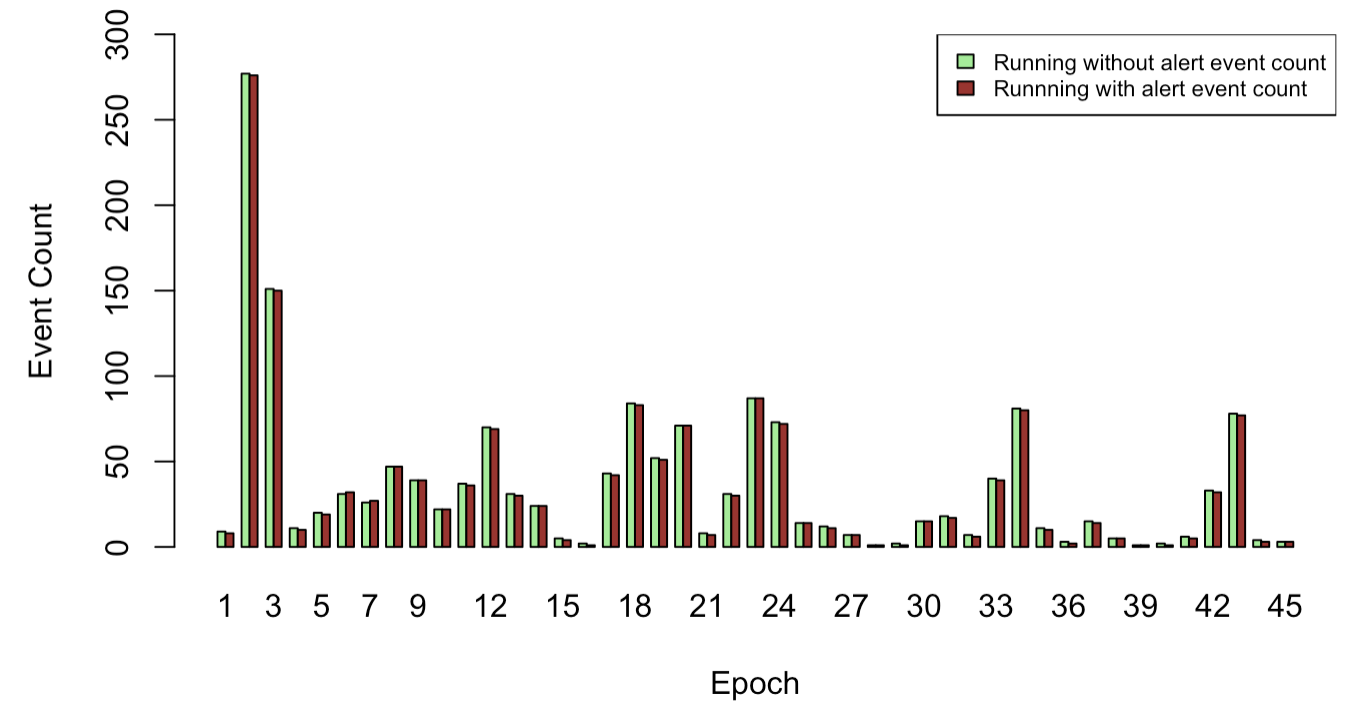}
	\caption{Frequency distribution of the number of two types of events (``running without alert'' and ``running with alert'') per epoch.}
	\label{fig5}
\end{figure}

Our dataset consists of 45 epochs, in which the total number of  ``running with alert'' events is $1584$, whereas the total number of  ``running without alert'' events is $1606$ (i.e., $3190$ running events and $45$ failures). Figure~\ref{fig5} presents epoch-wise counts of the two types of events. Since the two events occur alternatively, the height of two side-by-side bars are more or less the same. However, the number of events vary across  epochs, and hence the heights of bars vary with epoch. 

\begin{figure}[H]
	\centering
	\includegraphics[width=0.85\textwidth]{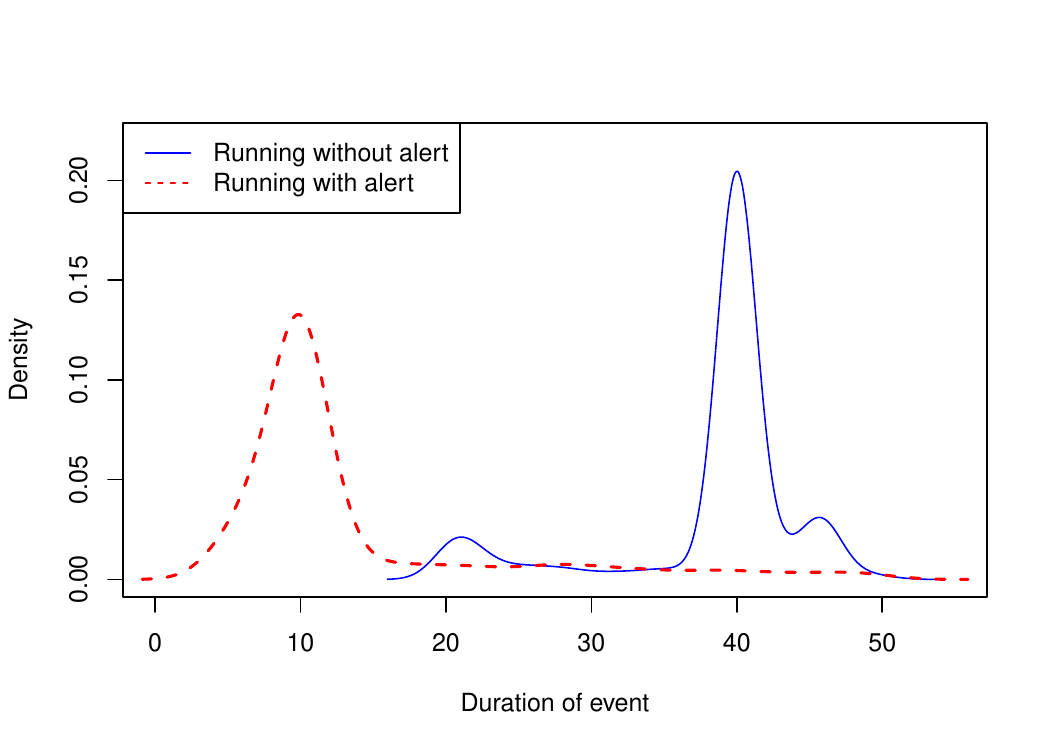}
	\caption{Distribution of the duration of the two types of events prior to failures.}
	\label{fig6}
\end{figure}

Figure~\ref{fig6} compares the distribution of ``running without alert'' event duration and ``running with alert'' event duration for all 3190 events in 45 epochs. It is clear that the average time spent on ``running without alert'' state is significantly more than the time spent on ``running with alert'' state. This supports the well-being of the machine.

To reiterate, the IM machine considered here consists of 72 different sensors that may explain the reasons behind the time spent on the three states. These sensors are majorly related to  \textit{mold surface temperature}, \textit{cooling rate of cavities}, \textit{post gate cavity pressure}, \textit{filled area of post gate cavity}, \textit{filled area of molding}, \textit{injection fill time}, \textit{screw runtime} etc. to name a few. Section~\ref{Modeling and analysis of sensor data} presents an elaborate discussion on these sensor based data.

\section{Model description and estimation}\label{model description and estimation}

In this section we propose a relational model between different types of events without using sensor-level data. This model along with the sensor data will be subsequently used in Section~\ref{Modeling and analysis of sensor data} for building the predictive model for determining the time to failure. First we discuss a few notations and distributional assumptions. 

\begin{longtable}{ L{2.0cm} | L{12.0cm} }
    \hline
    \textbf{\textit{Notation}} & \textbf{\textit{Description}}\\ \hline \endhead
    \hline
    $n$ :& Number of epochs observed. \\
    $N_{i}^1$ :& Number of ``running without alert'' events in the $i^{th}$ epoch; $i=1,\ldots,n$. \\
    $N_{i}^2$ :& Number of ``running with alert'' events in the $i^{th}$ epoch; $i=1,\ldots,n$. \\
    $R_i$ : & Total number of events in the $i^{th}$ epoch, i.e, $R_i=N_{i}^1+N_{i}^2$; $i=1,\ldots,n$. \\
    $N^1$ : & Total number of ``running without alert'' events in $n$ epochs; i.e. $N^1 = \sum_{i=1}^n N_{i}^1$.\\
    $N^2$ : & Total number of ``running with alert'' events in $n$ epochs; i.e. $N^2= \sum_{i=1}^n N_{i}^2$.\\
    	$N$ : & Total number of events in $n$ epochs; i.e, $N=\displaystyle\sum_{i=1}^n R_i = N^1 + N^2$.\\
    $X_{ij}^1$ :& Duration  of the $j^{th}$ ``running without alert'' event in the $i^{th}$ epoch; $j=1,\ldots,N_{i}^1$;~ $i=1,\ldots,n$. \\
    $X_{ij^{'}}^2$ :& Duration  of the $j^{'th}$ ``running with alert'' event in the $i^{th}$ epoch; $j^{'}=1,\ldots,N_{i}^2$;~ $i=1,\ldots,n$. \\
	\hline
\end{longtable}

In our data, $n=45,\ N^1=1606,\ N^2=1584$ and $N=3190$. The distribution of $N_i^1$ and $N_i^2$ are shown in Figure~\ref{fig5}, and the distribution of $X_{ij}^1$ and $X_{ij'}^2$ are depicted in Figure~\ref{fig6}. We make the following assumptions on the distribution of the associated random variables. For $j=1,\ldots,N_{i}^1$; $j^{'}=1,\ldots,N_{i}^2$ and $i=1,\ldots,n$,

\begin{itemize}
    \item $X^k_{ij}$ follow exponential distribution with mean $1/\lambda_k$, for $k=1,2$. 
       
    \item Since an epoch experiences at least one event before the ``failure", $R_i$ can take values in $\{1, 2, ...\}$, and therefore assumed to follow a shifted Poisson distribution with the probability mass function (PMF) given by
    \begin{equation}
    P(R_i=r_i) = e^{-\mu}~\frac{\mu^{(r_i-1)}}{(r_i-1)!}~;~r_i= 1, 2,\cdots.
    \label{eq:shifted-poisson}
	\end{equation}
\end{itemize}
Furthermore, all random variables, $X_{ij}^1,\ X_{ij^{'}}^2$, and $R_i$ are independent of each other for all $j=1,\ldots,N^1_i;~j^{'}=1,\ldots,N^2_i; ~i=1,\ldots,n$.

Although the distributional parameters may depend on the sensor-level data of the IM machine, we first form a base model using the event-level data (without sensor-level information) via likelihood approach. Recall that, in any given epoch  two events ``running without alert'' and ``running with alert''   occur alternatively (i.e., we cannot have two consecutive occurrences of, say, ``running without alert''), and an epoch can start with either of the two events. Suppose the choice of the first event in an epoch follows Bernoulli distribution with $p$ being the probability of the first event equal to ``running without alert''. Thus, the likelihood of event-level data for the $i$-th epoch can be written into four cases:

\begin{itemize}
\item \textbf{Situation 1}: The epoch starts with the event ``running without alert'' and the number of events $r_i$ is odd. Hence $N_i^1=\frac{r_i+1}{2}$ and $N_i^2=\frac{r_i-1}{2}$. 
\item \textbf{Situation 2}: The epoch starts with the event ``running without alert'' and the number of events $r_i$ is even. Hence $N_i^1=\frac{r_i}{2}$ and $N_i^2=\frac{r_i}{2}$.  

\item \textbf{Situation 3}: The epoch starts with the event ``running with alert'' and the number of events $r_i$ is odd. Hence $N_i^1=\frac{r_i-1}{2}$ and $N_i^2=\frac{r_i+1}{2}$.  

\item \textbf{Situation 4}: The epoch starts with the event ``running with alert'' and the number of events $r_i$ is even. Hence $N_i^1=\frac{r_i}{2}$ and $N_i^2=\frac{r_i}{2}$.  

\end{itemize}

Further assume that $S_l$ is the set of all epochs under the $l$-th situation, and let $|S_l|=n_l$ be the size of $S_l$, for $l=1,2,3,4$. Then, $n_1 + n_2+n_3+n_4 = 45$. Let  $\theta=(\lambda_1, \lambda_2, \mu, p)$ represent the set of distributional parameters to be estimated. The likelihood for Situation~1 can be written as: 
\begin{equation} \label{likelihood_situation_1}
L_1(\theta) = c_1 \displaystyle\prod_{i\in S_1}\left[P(R_i=r_i) \times p \times \displaystyle\prod_{j=1}^{\frac{r_{i}+1}{2}} f^{1}(x^1_{ij}) \times \displaystyle\prod_{j'=1}^{\frac{r_{i}-1}{2}} f^{2}(x^2_{ij'})\right ], 
\end{equation} 
where,  $f^k(.)$ is the probability density function (PDF) of exponential distribution with mean $1/\lambda_k$, for $k=1,2$ and $c_1$ is the proportionality constant independent of the parameters $\theta$. After ignoring the constant, using appropriate PDFs and PMF in the above likelihood function, and taking natural-log we get,
\begin{equation}\label{log_likelihood_situation_1}
\begin{aligned}
\mathcal{L}_1(\theta) = & \displaystyle -n_{1}~\mu + ln(\mu)\sum_{i\in S_{1}}(r_i-1) - \sum_{i\in S_{1}}ln((r_i-1)!)  + ln(\lambda_1)\sum_{i\in S_{1}}\biggl(\frac{r_{i}+1}{2}\biggr) \\ &
-\lambda_1 \displaystyle\sum_{i\in S_{1}}\sum_{j=1}^{\frac{r_{i}+1}{2}}x^1_{{i}j} + ln(\lambda_2)\sum_{i\in S_{1}}\biggl(\frac{r_{i}-1}{2}\biggr)  -\lambda_2 \sum_{i\in S_{1}}\sum_{j=1}^{\frac{r_{i}-1}{2}}x^2_{{i}j}+ n_1 ln(p) .  
\end{aligned}
\end{equation}

For other situations, the likelihood expression will be similar and have been discussed in Appendix~A1. Subsequently, the log-likelihood of the data from all $n$ epochs and four situations can be written as, 
\begin{equation}\label{log_likelihood}
    \mathcal{L}(\theta) = \mathcal{L}_1(\theta) +\mathcal{L}_2(\theta) +\mathcal{L}_3(\theta) +\mathcal{L}_4(\theta).
\end{equation}
The parameter vector $\theta=(\lambda_1, \lambda_2, p, \mu)$ is estimated by maximizing $\mathcal{L}(\theta)$ in Equation~(\ref{log_likelihood}). {Hence by defining $N_{il}^s=\frac{r_i+a_l^s}{2}$ and 
$$a_l^s=
\begin{cases}
(-1)^{s+1}, \text{~if $l=1$}\\
0, \text{~if $l=2,4$}\\
(-1)^s, \text{~if $l=3$}\\
\end{cases}
$$
for  $s=1,2$ and} from Equation~(\ref{log_likelihood}) one can find the closed form analytical expression of the  maximum likelihood estimators (MLEs) given by
\begin{flalign}\label{mles}
\widehat\lambda_s = \frac{\displaystyle\sum_{l=1}^4\sum_{i\in S_{l}}N_{il}^s}{\displaystyle\sum_{l=1}^{4}\sum_{i\in S_{l}}\sum_{j=1}^{N_{il}^s} x_{ij}^s}~ \text{for}\ s=1,2,~~ \widehat p = \frac{n_1+n_2}{n}\ \text{and}\ \widehat\mu = \frac{1}{n}\sum_{l=1}^{4}\sum_{i \in S_l}(r_{i}-1).
\end{flalign}  

Derivations of the above expressions are {presented in Appendix A2}. Although the parameters have closed form estimates, finding exact distribution, or at least variance of the estimators in Equation~(\ref{mles}), a requirement for computing the standard error (or equivalently, the confidence interval), is non-trivial. Therefore, we rely on asymptotic confidence intervals and parametric bootstrap-based confidence intervals for the parameters $\lambda_1, \lambda_2, p, \mu$.

Based on the asymptotic normality of the MLE of a parameter $\eta \in \{\lambda_1, \lambda_2, p, \mu\}$, a $100(1-\alpha)\%$ confidence interval for $\eta$ is 
\begin{equation*}\label{eq41}
\left[ \hat{\eta} - z_{\alpha/2}\frac{1}{\sqrt{I(\eta)}}, ~\hat{\eta} + z_{\alpha/2}\frac{1}{\sqrt{I(\eta)}}\right],
\end{equation*}
where $z_{\alpha/2}$ is the upper $\alpha/2$-quantile of the standard normal distribution, and ${I}(\eta)$ is the Fisher's information associated with the parameter $\eta$ defined as 
$${I}(\eta)=-E\left(\frac{\partial^2}{\partial \eta^2} \mathcal L(\theta)\right).$$
In practice the observed Fisher information $I(\hat{\eta})=-\frac{\partial^2}{\partial \eta^2} \mathcal L(\theta)$ evaluated at $\theta=\hat{\theta}$ can be used instead of ${I}(\eta)$. The observed Fisher information for our model parameters are as follows (derivation is straightforward and hence omitted):
$${I}(\hat{\lambda}_k)= \frac{\sum\limits_{l=1}^4\sum\limits_{i \in S_l}N^k_{il}}{\hat{\lambda}_k^2},\  \text{for}\ k=1,2,\ I(\hat{\mu})=\frac{n}{(\hat{\mu}+1)^2},\ \text{and}\  I(\hat{p})=\frac{n_1+n_2}{\hat{p}^2} + \frac{n_3+n_4}{(1-\hat{p})^2}.$$

Bootstrap confidence interval construction follows a resampling method. In this paper we adopt a parametric bootstrap approach. {One may refer to  \cite{Efron} for a book length reference.
The $100(1-\alpha)\%$ bootstrap confidence interval for any $\eta \in \{\lambda_1, \lambda_2, p, \mu\}$ is given by $$\big[\hat{\eta}_{([M\frac{\alpha}{2}])}, \hat{\eta}_{([M(\frac{1-\alpha}{2})])}\big],$$ where, $[x]$ represents the greatest integer of $x$, not exceeding $x$, and $\hat{\eta}_{(t)}$ denotes the $t$-th percentile of the bootstrap sample $\{\hat{\eta}_{m}, m = 1, 2, ..., M\}$. We have taken $M=2000$.
The key steps of the algorithm are summarized in Appendix A3.}

\section{Sensitivity analysis of the event-level model}\label{Analysis of data without covariates}
 
%
%
We first use the data on 3190 events to compute MLEs, asymptotic confidence intervals and bootstrap confidence intervals of the parameters, and then we perform a thorough simulation study to assess the sensitivity of the estimated parameters. Table~\ref{table_data_analysis_no_covar} includes the estimated MLE values, 95\% asymptotic confidence intervals and 95\% paramteric bootstrap confidence intervals of the model parameters.

\begin{table}[!h]\centering
\caption{MLEs and confidence intervals of parameters based on the base model.} 
	\begin{tabular}{|c|c|c|c|}\hline
{Parameters}  &{MLE}  & {{\thead{Asymptotic\\ Confidence Interval}}}        & {{\thead{Parametric Bootstrap\\ Confidence Interval}}} \\ \hline
$\lambda_1$         & 0.0261              & (0.0248, 0.0274)       & (0.0248, 0.0274)      \\
$\lambda_2$         & 0.0738             & (0.0712, 0.0774)       & (0.0702, 0.0776)      \\
$p$                 & 0.7778            & (0.6563, 0.8992)       & (0.6900, 0.8500)      \\
$\mu$       & 69.8889          & (67.4114, 72.3664)     & (66.4502, 71.3362)\\ 
\hline
\end{tabular}\label{table_data_analysis_no_covar}
\end{table}

A quick view of Table~\ref{table_data_analysis_no_covar} shows that $\lambda_1$ is significantly smaller than $\lambda_2$, which implies that the average time spent by the IM machine on ``running without alert" state is approximately three times (2.82 times to be more exact) more than the average time duration spent on ``running with alert". On average, each epoch consists of 70.89 events. Also, there is a 77\% chance that the first event in an epoch is ``running without alert".

{We now conduct a simulation study to assess the sensitivity of parameter estimates and robustness of the model.  The process starts with generating Monte Carlo samples from the assumed distributions and then estimate the parameters as per the formulae derived in Section~3. The model robustness is assessed by comparing the parameters used for data generation and the estimates obtained.} 

Since our dataset has 45 epochs, we start with the number of epochs close to  45 and then gradually increase it, i.e., the simulation study considers number of epochs equal to 50, 100 and 150. Keeping the observed ratio of $\lambda_2/\lambda_1$ in mind we take the values of $\lambda_1$ and $\lambda_2$ as 0.03 and 0.08, respectively as one set of values. We also take two different values for $\lambda_2$ as 0.03 and 0.02, which makes the ratio of $\lambda_2/\lambda_1$ as $1$ and $0.67$. The values of $\mu$ and $p$ are taken close to the MLEs obtained using the base model. The set of parameters used in the simulation study are:

\begin{itemize}
\item[(a)] $\theta_1= \{\lambda_1=0.03,~\lambda_2=0.08,~p=0.70~\text{and}~\mu=70\}$
\item[(b)] $\theta_2= \{\lambda_1=0.03,~\lambda_2=0.03,~p=0.67~\text{and}~\mu=68\}$
\item[(c)] $\theta_3=\{\lambda_1=0.03,~\lambda_2=0.02,~p=0.73~\text{and}~\mu=72\}.$ 
\end{itemize}

For each parameter combination, the data generation started with drawing $n$ (pre-specified number of epochs) $r_i$ values using the shifted Poisson distribution. The first event for each epoch is chosen as per the Bernoulli distribution with $p$ being the probability of observing ``running without alert". Subsequently, the duration of the two types of events were generated according to the exponential distributions with parameters $\lambda_1$ and $\lambda_2$, respectively.

For each set of parameters, we simulate 1000 datasets, and for each dataset, we followed the methodology discussed above to find the MLEs (as in Equation~(\ref{mles})). Let $\hat{\eta}_{i}$ denotes the estimate of $\eta \in \{\lambda_1, \lambda_2, p, \mu\}$ for the $i$-th dataset. Also let $L_{ai}$ and $U_{ai}$ respectively denote the lower limit and upper limit of 95\% asymptotic confidence interval, whereas $L_{bi}$ and $U_{bi}$ respectively denote the lower limit and upper limit of 95\% bootstrap confidence interval of $\eta$ (as discussed in Section~\ref{model description and estimation}). Here also, we used 2000 bootstrap samples for computing the 95\% bootstrap confidence intervals. Finally, the sensitivity of the parameters are assessed using the following four goodness of fit measures:

\begin{itemize}
\item Average bias:  
$$\frac{1}{1000}\sum\limits_{i=1}^{1000} |\hat{\eta}_{i}-\eta| .$$

\item Mean square error (MSE):  
$$\frac{1}{1000}\sum\limits_{i=1}^{1000} (\hat{\eta}_{i}-\eta)^2 .$$

\item Average 95\% asymptotic confidence interval: 
$$\left(\frac{1}{1000} \sum\limits_{i=1}^{1000} L_{ai},~ \frac{1}{1000} \sum\limits_{i=1}^{1000} U_{ai} \right).$$

\item Average 95\% bootstrap confidence interval: 
$$\left( \frac{1}{1000} \sum\limits_{i=1}^{1000} L_{bi},~ \frac{1}{1000} \sum\limits_{i=1}^{1000} U_{bi} \right).$$
\end{itemize}

\begin{figure}[!h]\centering
	\subfigure[Rate - ``running without alert" event]{
		\includegraphics[scale=0.65]{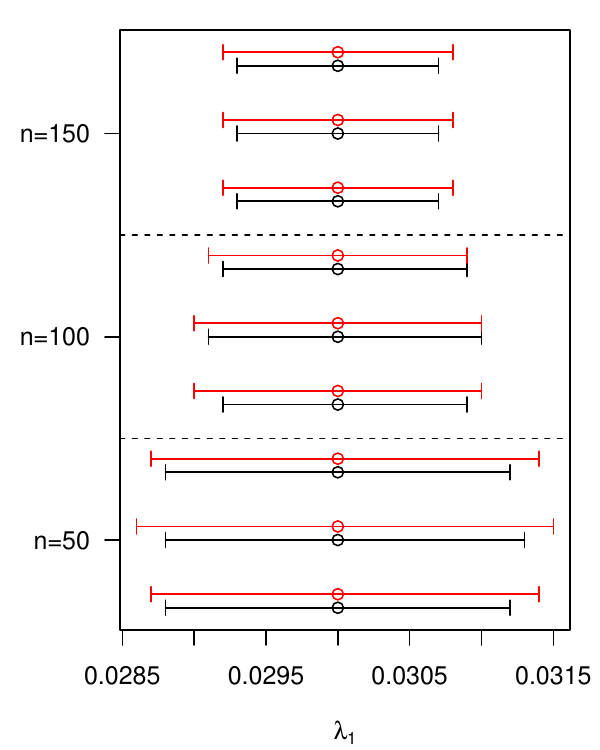}} 
	\subfigure[Rate - ``running with alert" event]{
		\includegraphics[scale=0.65]{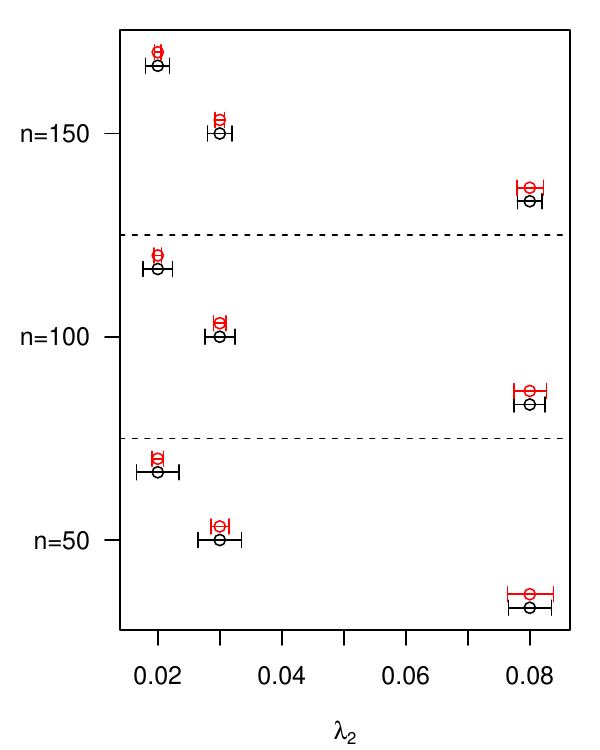}}
	\subfigure[Rate - number of events per epoch]{
	\includegraphics[scale=0.65]{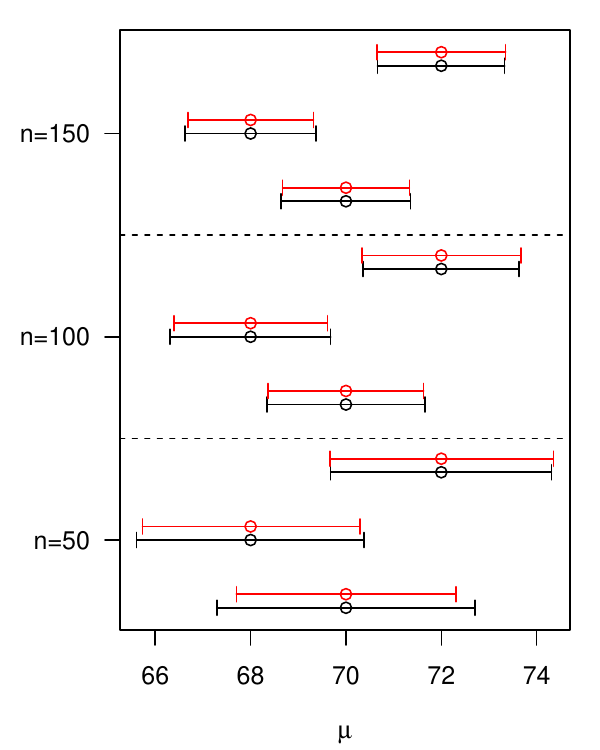}}
	\subfigure[P(epoch starts wth ``running without alert")]{
	\includegraphics[scale=0.65]{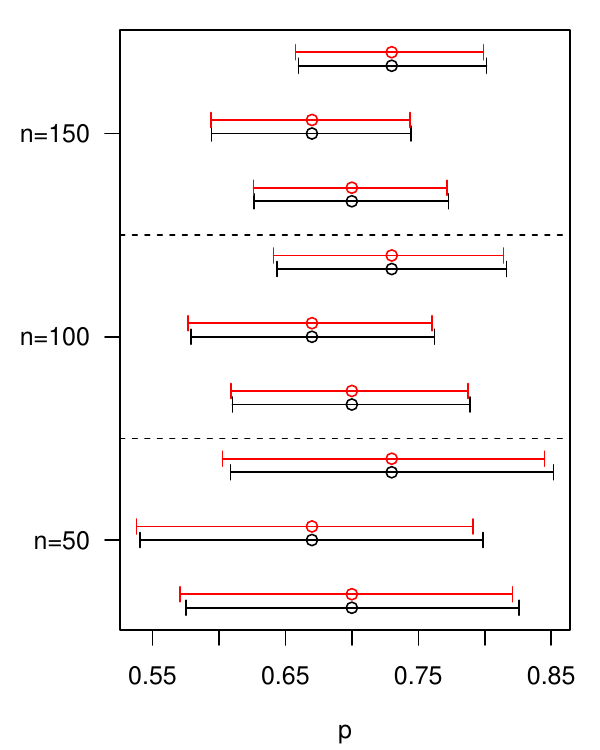}}
	\caption{Simulation based comparison of asymptotic (black lines) and bootstrap (red lines) confidence intervals for different parameters of the model. Each block separated by the dashed lines contains the results from three simulation study with true values: $\lambda_1=(0.03, 0.03, 0.03)$, $\lambda_2=(0.08, 0.03, 0.02)$, $\mu=(70,68,72)$ and $p=(0.70, 0.67, 0.73)$.}
	\label{Fig:param_simulation}
\end{figure}

The results are summarized in Tables~\ref{simulation_first} - \ref{simulation_third}.  For easier comparison, Figure~\ref{Fig:param_simulation} depicts the confidence intervals for all four parameters. Here, the red lines correspond to the bootstrap CIs whereas the black lines represent the asymptotic CIs.

\begin{table}[!h]
\centering
\caption {Simulation results under $\{\lambda_1 = 0.03, \lambda_2 = 0.08, p = 0.7, \mu = 70\}$ } \label{simulation_first}
{\begin{tabular}{|c|c|c|c|c|c|}\hline
{Parameter}  & {$n$} & {\thead{Average\\ Bias}} & {MSE}  & {{\thead{Average Asymptotic\\ Confidence Interval}}}        & {{\thead{Average  Bootstrap\\ Confidence Interval}}} \\ \hline
$\lambda_1$          & 50             &  0.0005       & 5.0713e-07    & (0.0288, 0.0312)              & (0.0287, 0.0314) \\
$\lambda_2$          & 50             &  0.0015       & 3.6271e-06    & (0.0766, 0.0835)              & (0.0764, 0.0838)\\
p                             & 50             &  0.0514       & 0.0042            & (0.5751, 0.8259)             & (0.5707, 0.8208)\\
$\mu$                     & 50             &  0.9429       & 1.3962            & (67.2930, 72.7170)         & (67.7036, 72.3167)\\ \hline
$\lambda_1$          & 100            &  0.0004       & 2.5451e-07    & (0.0292, 0.0309)             & (0.0290, 0.0310)\\
$\lambda_2$          & 100            &  0.0011       & 1.7817e-06     & (0.0775, 0.0825)             & (0.0775, 0.0827)\\
p                             & 100            &  0.0370       & 0.0021            & (0.6101,0.7888)              & (0.6092, 0.7875)\\
$\mu$                     & 100            &  0.6670       & 0.6988            & (68.3382, 71.6597)         & (68.3776, 71.6330)\\ \hline
$\lambda_1$          & 150            &  0.0003       & 1.6912e-07     & (0.0293, 0.0307)             & (0.0292, 0.0308)\\
$\lambda_2$          & 150            &  0.0009       & 1.1934e-06     & (0.0780, 0.0820)             & (0.0779, 0.0822)\\
p                             & 150            &  0.0296       & 0.0014            & (0.6265, 0.7727)             & (0.6259, 0.7714)\\
$\mu$                     & 150            &  0.5448       & 0.4664            & (68.6443, 71.3563)         & (68.6727, 71.3374)\\ \hline
\end{tabular}}
\end{table}

\begin{table}[!h]
\caption {Simulation results under $\{\lambda_1 = 0.03, \lambda_2 = 0.03, p = 0.67, \mu = 68\}$ } \label{simulation_second}
{\begin{tabular}{|c|c|c|c|c|c|} \hline
{Parameter}  & {$n$} & {\thead{Average\\ Bias}} & {MSE}  & {{\thead{Average Asymptotic\\ Confidence Interval}}}        & {{\thead{Average  Bootstrap\\ Confidence Interval}}} \\  \hline
$\lambda_1$          & 50             & 0.0006       & 5.2678e-07    & (0.0288, 0.0313)              & (0.0286, 0.0315 )  \\
$\lambda_2$          & 50             & 0.0006       & 5.2345e-07    & (0.0265, 0.0335)              & (0.0286, 0.0315)  \\
p                             & 50             & 0.0535       & 0.0044           & (0.5408, 0.7985)              & (0.5381, 0.7910)    \\
$\mu$                   & 50             & 0.9314       & 1.3631           & (65.6163, 70.3825)          & (65.7293, 70.2892)  \\
\hline
$\lambda_1$          & 100            & 0.0004      & 2.5451e-07    & (0.0291, 0.0310)              & (0.0290, 0.0310)  \\
$\lambda_2$          & 100            & 0.0004      & 2.6276e-07    & (0.0276, 0.0325)              & (0.0290, 0.0310)  \\
p                             & 100            & 0.0375      & 0.0022           & (0.5789, 0.7621)              & (0.5766, 0.7605)  \\
$\mu$                   & 100            & 0.6571      & 0.6796           & (66.3144, 69.6844)          & (66.3969, 69.6132)  \\
 \hline
$\lambda_1$          & 150            & 0.0003       & 1.7404e-07    & (0.0293, 0.0307)              & (0.0292, 0.0308)  \\  
$\lambda_2$          & 150            & 0.0003       & 1.7629e-07    & (0.0280, 0.0320)              & (0.0292, 0.0308)  \\
p                             & 150            & 0.0309       & 0.0015           & (0.5945, 0.7445)              & (0.5942, 0.7438)  \\
$\mu$                   & 150            & 0.5367       & 0.4534           & (66.6237, 69.3752)          & (66.6899, 69.3223) \\ \hline    
\end{tabular}}
\end{table}

\begin{table}[!h]
\centering
\caption {Simulation results under $\{\lambda_1 = 0.03, \lambda_2 = 0.02, p = 0.73, \mu = 72\}$ } \label{simulation_third}
{\begin{tabular}{|c|c|c|c|c|c|} \hline
{Parameter}  & {$n$} & {\thead{Average\\ Bias}} & {MSE}  & {{\thead{Average Asymptotic\\ Confidence Interval}}}        & {{\thead{Average  Bootstrap\\ Confidence Interval}}} \\   \hline
$\lambda_1$           & 50             & 0.0005       & 4.6586e-07   & (0.0288, 0.0312)         & (0.0287, 0.0314)     \\
$\lambda_2$           & 50             & 0.0004       & 2.2435e-07   & (0.0166, 0.0234)         & (0.0191, 0.0209)     \\
p                              & 50             & 0.0504       & 0.0039          & (0.6088, 0.8516)         & (0.6026, 0.8451)     \\
$\mu$                    & 50             & 0.9596       & 1.4462          & (69.6827, 74.3146)     & (69.6657, 74.3584)   \\
\hline
$\lambda_1$           & 100            & 0.0004       & 2.4750e-07   & (0.0292, 0.0309)         & (0.0291, 0.0309)     \\
$\lambda_2$           & 100            & 0.0003       & 1.0891e-07   & (0.0176, 0.0224)         & (0.0194, 0.0206)     \\
p                              & 100            & 0.0357       & 0.0020          & (0.6435, 0.8164)         & (0.6411, 0.8140)     \\
$\mu$                    & 100            & 0.6772       & 0.7209          & (70.3612, 73.6363)     & (70.3463, 73.6692)     \\
\hline
$\lambda_1$           & 150            & 0.0003       & 1.6373e-07   & (0.0293, 0.0307)         & (0.0292, 0.0308)     \\
$\lambda_2$           & 150            & 0.0002       & 7.4764e-08   & (0.0180, 0.0219)         & (0.0195, 0.0205)     \\
p                              & 150            & 0.0291       & 0.0013          & (0.6599, 0.8014)         & (0.6577, 0.7990)     \\
$\mu$                   & 150             & 0.5527       & 0.4801          & (70.6622, 73.3362)     & (70.6526, 73.3541)   \\
\hline
\end{tabular}}
\end{table}

From Tables~\ref{simulation_first} -- \ref{simulation_third}, and Figure~\ref{Fig:param_simulation}, some of the observations are very clear. In all these cases as epoch size increases, the performance of all estimators improve in terms of  biases, MSEs and the length of confidence intervals for most of the cases. It is also clear that the performance of confidence interval based on asymptotic and percentile bootstrap are quite satisfactory and they are quite close to each other in most of the cases. Interestingly, the length of average asymptotic confidence interval for $\lambda_2$ is more than the length of average bootstrap confidence interval, which differs from the trends on other parameter CIs.

\section{Modeling and analysis of sensor-level data}\label{Modeling and analysis of sensor data}
We now consider the modeling and analysis of all data in the presence of 72 different sensors operational in this IM machine. The dataset contains 3235 rows corresponding to 45 failure and 3190 running events (1584 ``running with alert" and 1606 ``running without alert"). It is expected that only a subset of these sensors might have useful information that can explain the current state of the machine, i.e., ``running without alert'', ``running with alert'' or ``failure''. We apply an efficient and reliable machine learning model called Random Forest (RF) for the identification of important features (sensors) that lead to accurate prediction of the current state of the machine. One can refer to \cite{ESL} for a detailed discussion on RF method.  

We follow a robust approach and build five different classifiers, and then find the union of top ten important features from these five classifiers. The five classifiers assume: 
\begin{itemize}
\item[(a)] M1 (3 classes): ``running without alert'', ``running with alert'' and ``failure'', 
\item[(b)] M2 (2 classes): “running   with alert” vs. “failure”, 
\item[(c)] M3 (2 classes): “running   without alert” vs. “failure”, 
\item[(d)] M4 (2 classes): ``running" (clubbing ``running with alert'' and ``running without alert'') vs. ``failure'', 
\item[(e)] M5 (2 classes): "running with alert" vs. "running without alert".  
\end{itemize}
The RF method was implemented by first splitting the full data with (3235 rows) in train and test as per 70:30 ratio. The performance of these models are summarized in Table \ref{classificatiom model cv}.

\begin{table}[h]\centering
\caption[Classification model performance summary and the important sensors]{Performance comparison of RF classifiers and top ten important sensors identified by the respective models.}
\label{classificatiom model cv}
\begin{tabular}{|m{1.75in}|m{0.5in}|m{0.5in}|m{1.75in}|}
\hline \multicolumn{1}{|c|}{{Model}} & \multicolumn{1}{c|}{{\thead{Training\\ Accuracy}}}	&	\multicolumn{1}{c|}{{\thead{Testing \\ Accuracy}}}	&	\multicolumn{1}{c|}{{Top 10 sensors}} \\ \hline
M1: "Running   With alert", "Running Without alert" and "Failure"  & 70.49\%                    & 52.61\%                   & $F_{22}$, $F_{24}$, $F_{27}$, $F_{29}$, $F_{30}$, $F_{31}$, $F_{32}$, $F_{34}$, $F_{67}$, $F_{68}$ \\ \hline
M2: “Running   with alert” vs. “Failure”                           & 98.24\%                    & 93.40\%                   & $F_{11}$, $F_{19}$, $F_{28}$, $F_{37}$, $F_{40}$, $F_{50}$, $F_{53}$, $F_{06}$, $F_{66}$, $F_{70}$ \\ \hline
M3: “Running   without alert” vs. “Failure”                        & 98.17\%                    & 92.11\%                   & $F_{11}$, $F_{19}$, $F_{28}$, $F_{37}$, $F_{50}$, $F_{52}$, $F_{54}$, $F_{06}$, $F_{66}$, $F_{70}$ \\ \hline
M4: “Running” vs. “Failure”                  & 99.44\%                    & 95.85\%                   & $F_{21}$, $F_{22}$, $F_{24}$, $F_{27}$, $F_{29}$, $F_{30}$, $F_{31}$, $F_{34}$, $F_{35}$, $F_{67}$ \\ \hline
M5: "Running with alert" and "Running without alert"               &  43.57\%                   &
59.98\%                   & $F_{15}$, $F_{16}$, $F_{27}$, $F_{35}$, $F_{04}$, $F_{44}$, $F_{05}$, $F_{58}$, $F_{66}$, $F_{07}$ \\ \hline
\end{tabular}
\end{table}

From Table~\ref{classificatiom model cv}, it is observed that some sensors are common in multiple models, whereas there are few sensors which are selected in only one model. It turns out that there are 22 different sensors that are selected across the five models. For readers understanding a brief description of each of these 22 sensors is given in Appendix~A6. 

\begin{figure}[H]\centering
	\subfigure[Sensor $F_4$]{
		\includegraphics[scale=0.65]{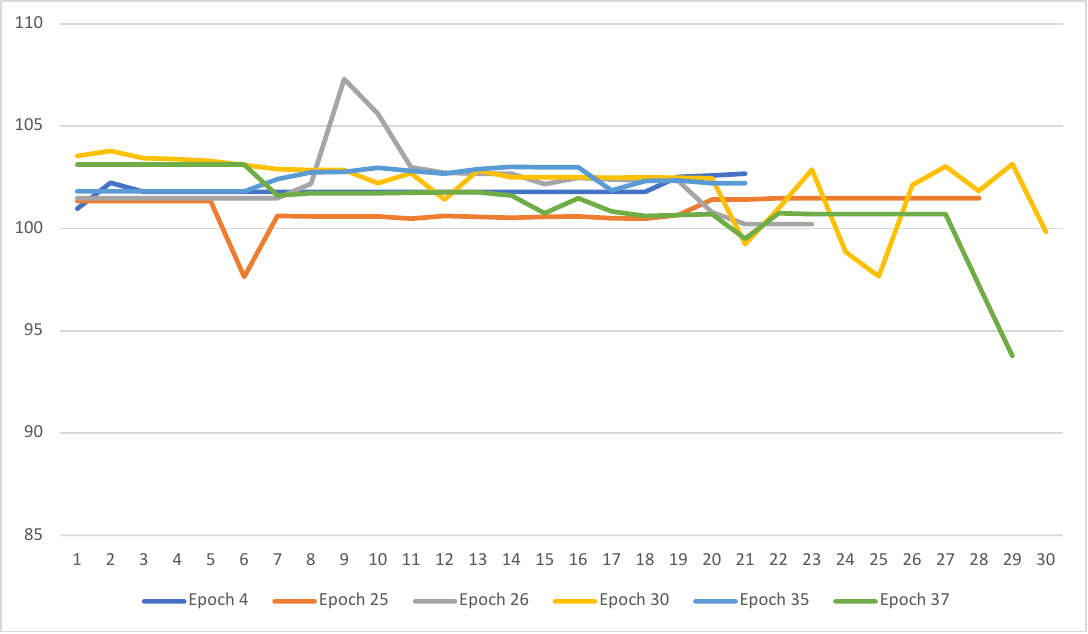}} 
	\subfigure[Sensor $F_{21}$]{
		\includegraphics[scale=0.65]{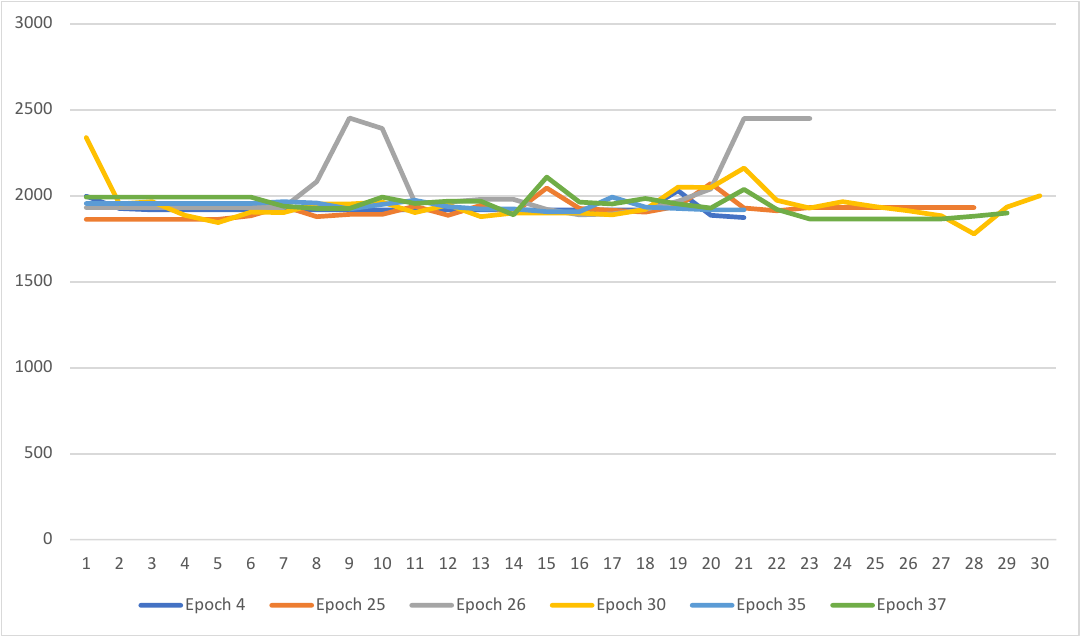}}
	\caption{Sensor data for $F_4$ and $F_{21}$ across epoch number: 4, 25, 26, 30, 35, 37.}
	\label{Fig:sensors_plot_first_two}
\end{figure}

To keep the notations simple let the $m(=22)$ sensors reported in Table~\ref{classificatiom model cv} (and in Table~\ref{description of selected cv} of Appendix~A6) are now renamed as $F_1, F_2, \cdots, F_m$. 
{To visualize the values of some of the sensors within and across some epochs, a plot of the values each of $F_4$ and  $F_{21}$ across the epochs  4, 25, 26, 30, 35, 37 is shown in Figure \ref{Fig:sensors_plot_first_two}. These epochs are chosen because the number of events in these epochs are comparable.} 

{The observed values of sensor $F_4$ are very close to each other in Epoch 4 whereas in Epoch 35, these values are little varying in its middle than the other sides. Towards the end of Epoch 30, the values are varying a lot compared to the other values in the same epoch. In Epoch 25 the 6th observation of $F_4$ is significantly lower than the other values although these values have been a little more after 20th observation. The sensor values from 8th observation to 12th observation in Epoch 26 are higher than the other values whereas in Epoch 37, the values are more or less same till 27th observation post which these values have been reduced significantly. The values of the sensor vary from little lower than 95 to little more than 106 across the epochs. 
The observed values of sensor $F_{21}$ are very close to each other in all the epochs shown. However, there is some variation in the values in Epoch 26. The values of the sensor vary from around 1777 to nearly 2500 across the epochs.
}

We now use these 22 sensor level data to build a predictive model for predicting the time spent by the machine on different states. For reducing the computation burden, we have modelled $\lambda_1, \lambda_2$ and $\mu$ with respect to the sensors, and used the MLE of $p$ (=0.7778) obtained from the base model (discussed in Sections~\ref{model description and estimation} and \ref{Analysis of data without covariates}). The generalized linear model (GLM) considered for  $\lambda_1, \lambda_2$ and $\mu$ in $i$-th epoch can be written as,
\begin{equation}\label{eq:glm-params}
\lambda_{1i}= \exp\left(\beta_0+\sum_{k=1}^m F_{ki} \beta_k\right), ~ \lambda_{2i}= \exp\left(\gamma_0+\sum_{k=1}^m F_{ki} \gamma_k\right),\ \mu_i= \exp\left(\eta_0+\sum_{k=1}^m F_{ki} \eta_k\right),
\end{equation}
where $\beta_k, \gamma_k, \eta_k$ for $k=0,1,\cdots, m$ denote the unknown regression coefficients and $F_{ki}$ denotes the $F_k$-th sensor value in the $i$-th epoch.

Recall from Section~\ref{model description and estimation}, that the log-likelihood of the data is $\mathcal{L}(\theta)= \mathcal{L}_1(\theta) + \mathcal{L}_2(\theta) + \mathcal{L}_3(\theta) + \mathcal{L}_4(\theta)$, where $\mathcal{L}_l(\theta)$ represents the likelihood under Situation~$l$, based on whether the first event is ``running without alert" and the number of events in an epoch is odd or even. Moreover, $\mathcal{L}_l(\theta)$ is an additive separable function with respect to the parameters in $\{\lambda_1, \lambda_2, \mu, p\}$. If the regression parameters are denoted by $\beta=(\beta_0, \beta_1, \cdots, \beta_m)$, $\gamma=(\gamma_0, \gamma_1, \cdots, \gamma_m)$ and $\eta=(\eta_0, \eta_1, \cdots, \eta_m)$,  then  
\begin{equation}\label{Situation s log likelihood glm}
\mathcal{L}_l(\theta)=\mathcal{L}_l(\beta, \gamma, \eta, p)= g_{1l}(\beta) + g_{2l}(\gamma) + g_{3l}(\eta) +g_{4l}(p), 
\end{equation}
where, $g_{1l}(\beta)$, $g_{2l}(\gamma)$, $g_{3l}(\eta)$ and $g_{4l}(p)$  are  functions involving $\beta$, $\gamma$, $\eta$  and $p$ only. Refer to Appendix~A4 for the expression of $g_{1l}(\beta)$, $g_{2l}(\gamma)$ and $g_{3l}(\eta)$. Since we are using the MLE of $p$ obtained in Section~\ref{Analysis of data without covariates}, the goal is to estimate the parameters $\eta, \beta, \gamma$ by maximizing the function $\sum_{l=1}^4 \mathcal{L}_l(\eta, \beta, \gamma, \hat{p})$. Due to separability of the terms in log-likelihood, the parameters are estimated by solving three separate maximization problem, that is, maximize $\sum_{l=1}^4 g_{1l}(\beta)$, $\sum_{l=1}^4 g_{2l}(\gamma)$ amd $\sum_{l=1}^4 g_{3l}(\eta)$  respectively, for estimating $\beta, \gamma$ and $\eta$ vectors. The maximization of the above functions can be carried out by using any standard scientific computing software. We have used a built-in ``optim'' function in R software to estimate the parameters.

We further investigate the statistical significance of these 22 sensors with respect to this regression models for $\eta, \beta$ and $\gamma$ using a forward selection method coupled with Akaike information criteria (AIC). See \cite{Montgomery} for details on AIC and stepwise variable selection method. The forward selection method starts with no sensor in the model and subsequently adds sensors one by one. The final set of sensors is the set of sensors that gives minimum value of AIC of the associated model. The final set of statistically significant sensors selected to model $\lambda_1, \lambda_2$ and $\mu$ (corresponding to $\eta, \beta$ and $\gamma$ respectively) are listed in Tables~\ref{running without issue cv}, \ref{running with issue cv} and \ref{GLM for mu}.

%
%

\begin{table}[h]
\centering
\begin{longtable}{|l|l|}
\caption[Sensors selected to model $\lambda_1$]{Significant sensors identified by AIC criterion selected to model $\lambda_1$.} \label{running without issue cv} \\

\hline \multicolumn{1}{|c|}{\textbf{Sensors}} & \multicolumn{1}{c|}{\textbf{\thead{Estimate\\ of parameter}}} \\ \hline 
\endfirsthead

Intercept & 5.8208E-05 \\
Cooling Rate PostGate Pressure   drop Cavity 4 (per sec)   & 4.7690E-05 \\
Cooling Rate PostGate Pressure   drop Cavity 7 (per sec)   & -8.8394E-07 \\
Cooling Rate PostGate Pressure   drop Cavity 11 (per sec)     & 2.0106E-05  \\
Injection Fill Time(Sec)                             & 1.9799E-05\\
Injection Integral From Molding Machine				&	-1.8976E-04 \\ 
Mold Surface Temp Cavity 1                         & 3.4295E-05  \\
Mold Surface Temp Cavity 6                         & 1.6844E-05  \\
Mold Surface Temp Cavity 10                         & 1.7863E-05  \\
Mold Surface Temp Cavity 12                         & 4.4228E-05  \\ \hline
\end{longtable}
\end{table}

%
%

\begin{table}[h]
\centering
\begin{longtable}{|l|l|}
\caption[Sensors selected to model $\lambda_2$]{Significant sensors identified by AIC criterion selected to model $\lambda_2$.} \label{running with issue cv} \\

\hline \multicolumn{1}{|c|}{\textbf{Sensors}} & \multicolumn{1}{c|}{\textbf{\thead{Estimate\\ of parameter}}} \\ \hline 
\endfirsthead

%
%
Intercept                         & 6.0770E-05         \\
Cooling Rate PostGate Pressure   drop Cavity 4 (per sec) & 2.9760E-05         \\
Cooling Rate PostGate Pressure   drop Cavity 6 (per sec) & 9.1680E-05         \\
Cooling Rate PostGate Pressure   drop Cavity 7 (per sec) & 1.0137E-04       \\
Injection Fill Time(Sec)                & -5.9380E-05       \\
Mold Surface Temp Cavity 1             & 1.1336E-04         \\
Mold Surface Temp Cavity 5              & -1.9830E-05        \\
Mold Surface Temp Cavity 6              & 1.6074E-04         \\
Mold Surface Temp Cavity 10          & -1.3300E-06        \\
Mold Surface Temp Cavity 12             & 8.7760E-05         \\
Peak PostGate Cavity 15                 & -5.6669E-04       \\ \hline
\end{longtable}
\end{table}

\begin{table}[h]
\centering
\begin{longtable}{|l|l|}
\caption[Sensors selected to model $\mu$]{Significant sensors identified by AIC criterion  selected to model $\mu$.} \label{GLM for mu} \\

\hline \multicolumn{1}{|c|}{\textbf{Sensors}} & \multicolumn{1}{c|}{\textbf{\thead{Estimate\\ of parameter}}} \\ \hline  
\endfirsthead

%
%
Intercept                                				 & 8.0257E-00             \\
Cooling Rate PostGate Pressure drop Cavity 2 (per sec)          & 9.3773E-03             \\
Cooling Rate PostGate Pressure drop Cavity 4 (per sec)          & 1.6072E-03            \\
Cooling Rate PostGate Pressure drop Cavity 5 (per sec)			&	3.3294E-03	\\

Cooling Rate PostGate Pressure drop Cavity 6 (per sec)			&	-2.7174E-04	\\

Cooling Rate PostGate Pressure drop Cavity 7 (per sec)			&	5.67E-03	\\

Cooling Rate PostGate Pressure drop Cavity 9 (per sec)			&	-1.5940E-03	\\

Cooling Rate PostGate Pressure drop Cavity 10 (per sec)        & 9.8359E-04             \\
Cooling Rate PostGate Pressure drop Cavity 11 (per sec)        & 1.3243E-03             \\
Cooling Rate PostGate Pressure drop Cavity 13 (per sec)        & -5.5745E-03            \\
Cooling Rate PostGate Pressure drop Cavity 15 (per sec)			&	3.9870E-03	\\


Injection Integral PostGate Cavity 9                        & 4.9999E-03         \\

Injection Integral PostGate Cavity 15                       & -2.3998E-02         \\

Injection Integral PostGate Cavity 16                        & 6.2386E-04         \\

Injection Fill   Time(Sec)                                 & 15.2131E-00         \\ 

Injection Integral From Molding Machine						&	9.2841E-03		\\
Mold Surface Temp   Cavity 1                                    & 1.6174E-00     \\
Mold Surface Temp   Cavity 5                             & 4.3807E-00             \\
Mold Surface Temp   Cavity 6                             & -3.7911E-00            \\
Mold Surface Temp   Cavity 10                                 & 3.9269E-00     \\
Mold Surface Temp   Cavity 12                            & -6.4066E-00     \\

Peak PostGate Cavity 3                                         & 1.8124E-02     \\


Peak PostGate Cavity 15                                     & 7.1845E-02  \\ \hline
\end{longtable}
\end{table}

Clearly these statistically significant sensors are specifically responsible for ``running without alert'' events, ``running with alert'' events and average number of events occurring in an epoch. Some of these parameters from $\beta$ and $\gamma$ are with different signs in ``running without alert'' and ``running with alert'' events , for example Mold Surface Temp Cavity 10, Cooling Rate PostGate Pressure   drop Cavity 7 (per sec) etc. Also there are a few sensors with same sign in both of them. It is to be noted that the sensors are selected to comply with the fact that if running time without alert becomes more  in an epoch then running time with alert will be less in the same epoch and vice-versa. It is recommended to keep a close eye on all the chosen sensors given
by the model and adjust them appropriately so that the machine runs less frequently with alerts. As a result, the running time will increase and the failure rate will drop.

\section{Prediction of time to fail}\label{Prediction of Time to Failure}

We now predict the time to fail for this IM machine while in running stage. The predicted or expected time to fail can be expressed as $E[\text{Time to fail}]=E_R\left(E[\text{Time to fail}~ |~ R=r]\right)$, where $R$ is the number of events in an epoch,  which has been assumed to follow a shifted Poisson distribution (as in Equation (\ref{eq:shifted-poisson}). Given that we categorize the epochs in four situations (Section~\ref{model description and estimation}) based on whether the number of events is even or odd, and whether the first event is of ``running with alert" or ``running without alert",  the expected time to fail can be further broken down as
\begin{eqnarray}
E[\text{Time to fail}]&=&E_R\left(E[\text{Time to fail} | R=r]\right) 
=\sum_{r=1}^{\infty} E[\text{Time to fail} |R=r] P(R=r) \nonumber\\
&=& \sum_{r\in \{1,3,5,\cdots\}} E[\text{Time to fail} |R=r] P(R=r) \nonumber\\
&&+  \sum_{r\in \{2,4,6,\cdots\}} E[\text{Time to fail} | R=r] P(R=r)\nonumber  \\ 
&=& \frac{(1-e^{-2\mu})}{4} \left(\frac{\mu+1}{\lambda_1}+\frac{\mu+1}{\lambda_2}\right)\nonumber\\
&&+\frac{(1+e^{-2\mu})}{4} \Big[\frac{\mu+2p}{\lambda_1}+\frac{\mu+2(1-p)}{\lambda_2}\Big]. 
  \label{eq:prediction}
\end{eqnarray}
Derivation of the final expression in Equation~(\ref{eq:prediction}) is provided in  Appendix~A7. The evaluation of $E(\text{Time to fail})$ requires the estimation  of $\lambda_1, \lambda_2, \mu$ and $p$. As discussed in Section~\ref{Modeling and analysis of sensor data}, $\lambda_1$, $\lambda_2$ and $\mu$ depend on the sensor record data via the generalized linear regression model (in Equation~(\ref{eq:glm-params})), whereas, $p$ has an overall estimate of $\hat{p} =0.7778$.



We now compute the expected time to fail for each epoch from our IM machine dataset using the proposed model. For an illustration purpose we discuss the computation of the expected time to fail for the first epoch. 
It turns out that the observed data of the first epoch consists of 17 events: 9 of the type ``running without alert" (corresponding to $\lambda_1$), and 8 of the type ``running with alert" (represents $\lambda_2$). As per Table~\ref{running without issue cv}, there are nine significant sensors viz. $F_4$, $F_5$, $F_7$, $F_{16}$, $F_{22}$, $F_{29}$, $F_{32}$, $F_{68}$, $F_{69}$, which are used to model $\lambda_1$. The prediction formula for the expected time to fail requires one set of sensor values. We take the average of the observed values of the corresponding sensor. For instance, the value of $F_4$ (with respect to $\lambda_1$ estimation), is the average of 101.1783, 99.9445, 99.7284, 97.3887, 101.1746, 100.8110, 99.7719, 103.5241, 103.8081 ($F_4$ values for the nine events in Epoch~1). The averages for other sensor values were also computed using the nine events of this epoch and subsequently Equation (\ref{eq:glm-params}) was used to estimate $\lambda_1$. Similarly the representative values of 10 sensors (Table~\ref{running with issue cv}) from the 8 ``running with alert" events were used for the estimation of $\lambda_2$, and 22 sensor (Table~\ref{GLM for mu}) based values from 17 events were used to estimate $\mu$. The epoch-wise expected time to fail are reported in Table \ref{epoch_wise_time}.



\begin{table}[h]
\centering
\begin{longtable}{|l|l|l|l|l|l|}
\caption[Epoch-wise estimated time to failure]{Epoch-wise estimated time to failure.} \label{epoch_wise_time} \\

\hline \multicolumn{1}{|c|}{\textbf{Epoch}} & \multicolumn{1}{c|}{\textbf{\thead{Expected\\ time to fail}}} & \multicolumn{1}{|c|}{\textbf{Epoch}} & \multicolumn{1}{c|}{\textbf{\thead{Expected\\ time to fail}}} & \multicolumn{1}{|c|}{\textbf{Epoch}} & \multicolumn{1}{c|}{\textbf{\thead{Expected\\ time to fail}}} \\ \hline  
\endfirsthead

1	&	2590.47	&	16	&	35.02	&	31	&	425.27	\\
2	&	11407.73	&	17	&	3345.12	&	32	&	429.86	\\
3	&	6261.52	&	18	&	2951.75	&	33	&	2451.67	\\
4	&	232.38	&	19	&	3102.31	&	34	&	3331.50	\\
5	&	928.50	&	20	&	951.87	&	35	&	516.83	\\
6	&	3931.49	&	21	&	602.86	&	36	&	239.37	\\
7	&	1394.65	&	22	&	3569.43	&	37	&	2039.87	\\
8	&	3202.23	&	23	&	2833.86	&	38	&	49.27	\\
9	&	2555.12	&	24	&	1548.36	&	39	&	130.81	\\
10	&	1963.56	&	25	&	1357.88	&	40	&	108.31	\\
11	&	2680.50	&	26	&	1340.19	&	41	&	676.48	\\
12	&	3409.98	&	27	&	1161.21	&	42	&	2168.00	\\
13	&	1771.74	&	28	&	142.11	&	43	&	1855.81	\\
14	&	765.55	&	29	&	142.11	&	44	&	170.77	\\
15	&	869.05	&	30	&	1508.72	&	45	&	170.77	\\
	\hline
\end{longtable}
\end{table}

\begin{figure}[H]
	\centering
\includegraphics[width=0.95 \textwidth]{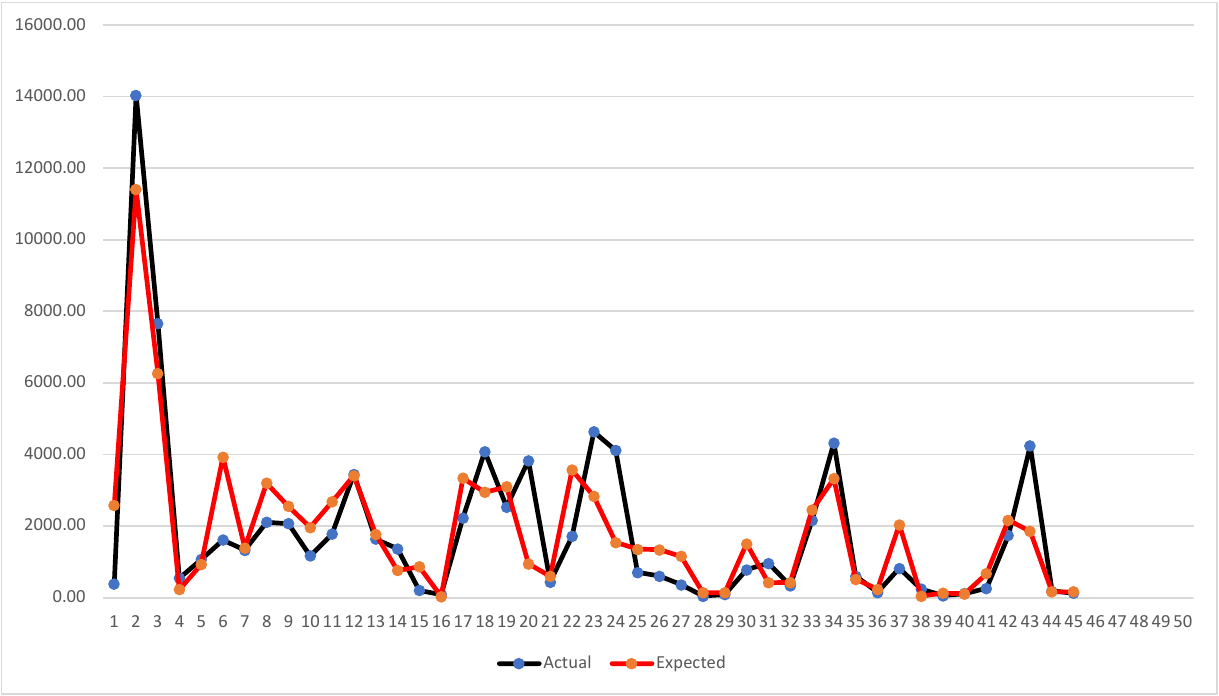}
	\caption{Plot of epoch-wise actual and expected time to fail by the proposed model.}
	\label{fig actual vs expected}
\end{figure}

Figure~\ref{fig actual vs expected} compares the actual and expected time to fail using the proposed model for each of the 45 epochs. From Figure~\ref{fig actual vs expected}, it is observed that expected time to failure and observed time to failure deviate by some amount for a few epochs. This is because of the fact that we have considered the average value of records for each sensor. 

\begin{figure}[H]\centering
		\includegraphics[scale=0.7]{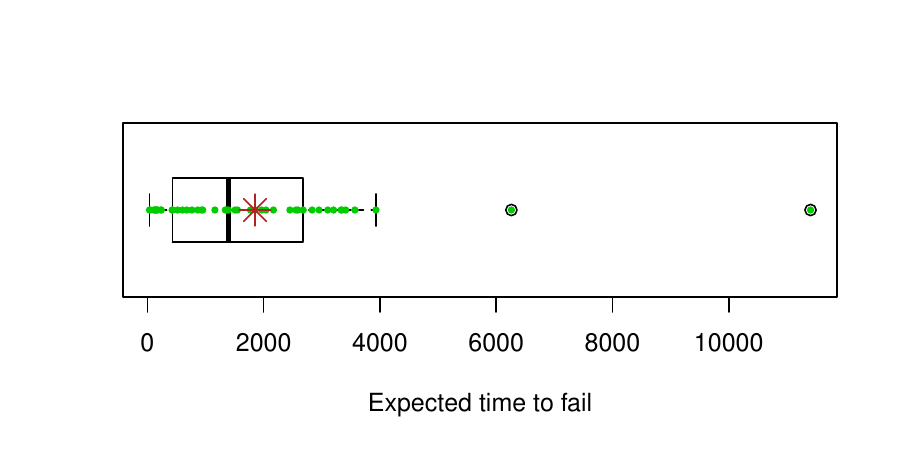}
	\caption{Boxplot and green dots show the distribution of the epoch-wise expected time to fail whereas the red asterisk represents an out-of-sample epoch prediction.}
	\label{Fig:Box_plot_expected_time_GLM}
\end{figure}

To compute the expected time to fail for an out-of-sample epoch, we have taken the average value of estimated time to fail from Table~\ref{epoch_wise_time} of all the 45 epochs. Figure~\ref{Fig:Box_plot_expected_time_GLM} presents a distribution of the expected time to fail for the 45 epochs and an out-of-sample epoch in the same plot using our proposed model. Alternatively, one can take the average sensor values across all 45 epochs (i.e., over 3190 events) to estimate the expected time to fail. Of course, if we know the true values of the sensors, one can use that instead, however these values are typically not known in advance.


\begin{figure}[H]\centering
	\includegraphics[scale=0.7]{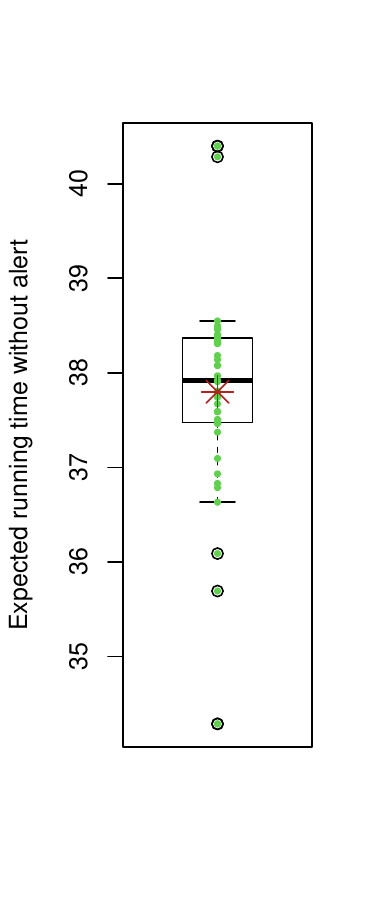}
	\includegraphics[scale=0.7]{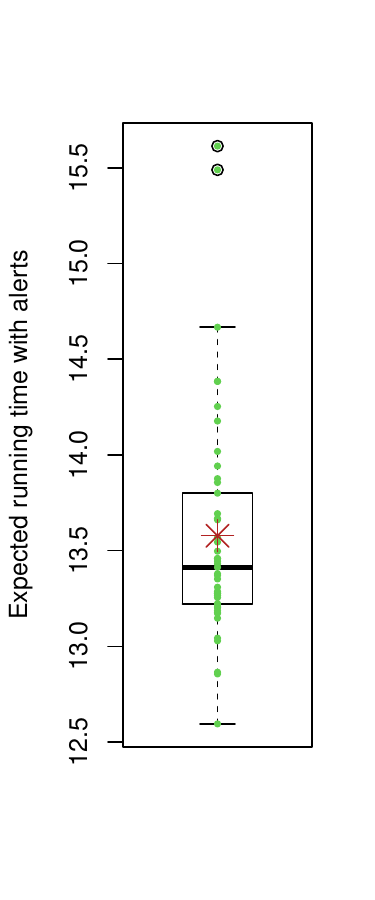}
	\includegraphics[scale=0.7]{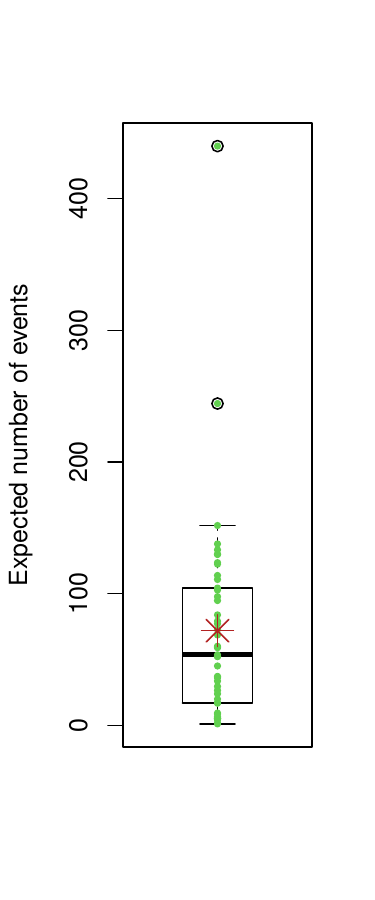}
  %
	\caption{Boxplots and green dots show the distribution of each of epoch-wise expected running time without alert, expected running time with alert, expected number of events whereas the red asterisk represent their average values.}
	\label{Fig:Box_plot_expected_and_out_of_sample}
\end{figure}

Additionally, we compare separate distribution of expected time spent on ``running without alert" and ``running with alert". For instance, the predicted expected time spent on ``running  without alert" in Epoch 1 is $1/\lambda_1$ which is estimated using only 9 (out of 17) events used for modeling $\lambda_1$. The left most panel of Figure~\ref{Fig:Box_plot_expected_and_out_of_sample} depicts the distribution.  Similarly, the middle panel shows epoch-wise distribution of the expected time spent on ``running with alert". The right most panel presents the expected value of $\mu+1$ from each of the 45 epochs. Figure~\ref{Fig:Box_plot_expected_and_out_of_sample} also shows that time duration spent on ``running without alert" is substantially longer than the time spent on ``running with alert", which is a good sign for a healthy machine.

\begin{figure}[H]
	\centering
\includegraphics[width=0.95 \textwidth]{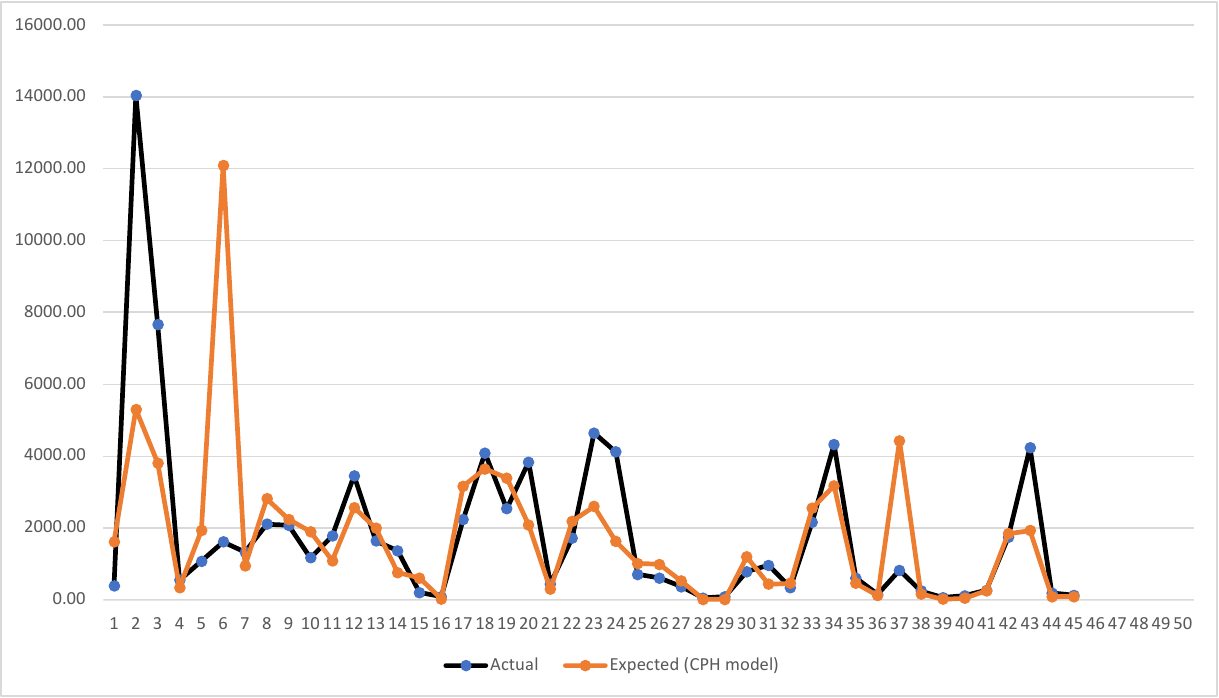}
	\caption{Plot of epoch-wise actual and expected time to fail by the CPH model.}
	\label{fig actual vs cph}
\end{figure}

Finally, we compare the performance of the proposed model with the popular Cox proportional hazard (CPH) model for predicting the expected time to fail. 
For a fair comparison, we have used the same 22 most significant sensors (presented in Table~\ref{classificatiom model cv}) which were selected using the random forest (RF) method. Figure~\ref{fig actual vs cph} depicts the actual and expected time to fail using CPH model, and the proposed model for the 45 epochs. 

\begin{figure}[H]
	\centering
\includegraphics[width=0.95 \textwidth]{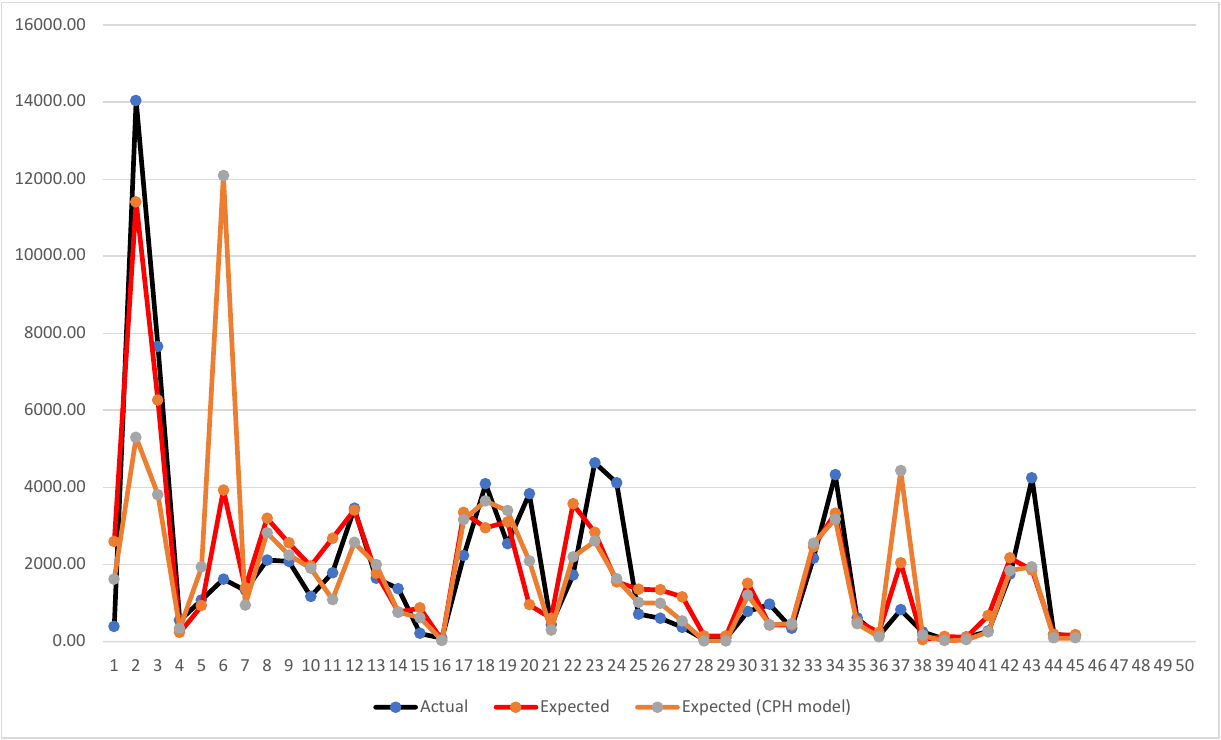}
	\caption{Plot of epoch-wise actual and expected time to fail by the proposed model and CPH model.}
	\label{fig actual vs expected vs cph}
\end{figure}

Figure~\ref{fig actual vs cph} shows that the actual and expected time to fail by using our proposed model are quite close to each other as compared to the CPH model, for instance, at Epochs 2, 3 and 6. For an easy reference we also have plotted actual time to fail and expected time to fail by using our proposed model and CPH model in Figure \ref{fig actual vs expected vs cph}.  It shows that our proposed model gives a better expected time to fail than that of given by CPH model.

{An overall comparison can also be made by computing the MSE, mean absolute error (MAE), maximum error (MaxE) and correlation between actual data and the proposed model vs correlation between actual data and the CPH model. These values are reported in Table \ref{Table:proposed_vs_cph}}

\begin{table}[h]\centering
{
\caption{{Performance of the proposed model vs CPH model}}
\label{Table:proposed_vs_cph}
\begin{tabular}{|c|c|c|c|c|}
\hline 
Model	&	MSE	&	MAE	&	MaxE	&	Correlation \\ \hline
Proposed	&	1299394.36	&	796.38	&	2881.12	&	0.89\\
CPH	&	5389542.03	&	1102.77	&	10477.12	&	0.48\\
\hline
\end{tabular}}
\end{table}

{From the above table, it is clear that the proposed model is a much better model than the CPH model. The MSE of the CPH model is more than 4 times than that of the proposed model. Also MAE and MaxE of the proposed model are much lower than those of CPH model. Finally the correlation between the actual failure time and the predicted failure time by the proposed model is much higher than the correlation between the actual failure time and the predicted failure time by the CPH model. Hence the proposed model can be used to predict the failure time of IM machines.}

\section{Conclusion}\label{Conclusion}

In this study we considered parametric analysis of the sequence of events data observed from an IM machine. The events in an epoch can be classified into ``running without alert'' and ``running with alert''  followed by failure of the machine. {The  ``running without alert'' must follow (or be followed by) ``running with alert''. Hence the data set is quite complex in nature and therefore it would be more appealing to fit a  model suitable more for the practitioner's point of view.}  For the sake of simplicity we assumed that  random variables associated with the duration for ``running without alert'' and ``running with alert''  follow one parameter exponential distribution and they are independent to each other. As a base model we started with a structure of the data when the presence of the sensors were not considered in the model.  The associated parameters were estimated using maximum likelihood estimation method.  Also 95\% asymptotic confidence interval and parametric bootstrap confidence interval of the parameters were reported.  

We analyzed the data further after taking the presence of the sensors into consideration. Since the number of sensors (or covariates) is large (72), we used random forest model for identifying the subset of significant sensors.  Forward selection method along with AIC was used for model ranking.  We discussed the maximum likelihood estimation of the parameters using a non linear dependency between the model parameters and the sensors present in the data.  Using the proposed model we compute the expected time to failure of the machine. It turns out that our model works better than the well known CPH model which justifies the usefulness of the current model.

Although we have assumed exponential distribution for modeling the time spent on different states like ``running without alert" or ``running with alert", one can alternatively look for a more general and widely applicable distribution like Weibull distribution or any other suitable distribution(s). In this paper, we have clubbed all alerts messages into one group and labelled them as ``running with alert", but it may be more insightful to further split this set of alerts into at least two groups, say, ``running with riskier alerts" and ``running with harmless alerts". This might be helpful in narrowing down the reasons behind the system failure. Subsequently, a competing risk approach based methodology may also be investigated for prediction of the expected time to fail. 
 \\

\section*{Acknowledgements}

The authors would like to thank the Editor and two referees for their helpful comments and suggestions that led to significant improvement in the manuscript.

\section*{Disclosure Statement}
No potential conflict of interest was reported by the
author(s).

\section*{Funding}
No funding was received. 

\section*{Data Availability Statement}
The data that support the findings of this study are available from the authors upon reasonable request.


\section*{Appendix~A1: Likelihood of the base model}

Details of the log-likelihoods $\mathcal{L}_1(\theta)$, $\mathcal{L}_2(\theta)$, $\mathcal{L}_3(\theta)$ and $\mathcal{L}_4(\theta)$ mentioned in Section~\ref{model description and estimation} are presented as follows.

\noindent \textbf{Situation 1}: The epoch starts with the event ``running without alert'' and the number of events $r_i$ is odd. The likelihood function  is 
\begin{equation*} 
    L_1(\theta) = c_1 \displaystyle\prod_{i\in S_1}\biggl [P(R_i=r_i) ~ p ~ \displaystyle\prod_{j=1}^{\frac{r_{i}+1}{2}} f^{1}(x^1_{ij}) ~ \displaystyle\prod_{j=1}^{\frac{r_{i}-1}{2}} f^{2}(x^2_{ij})\biggr ], 
\end{equation*}  

where,  $f^k(.)$ is the probability density function (PDF) of exponential distribution with mean $1/\lambda_k$, for $k=1,2$ and $c_1$ is the proportionality constant independent of the parameters $\theta$. After ignoring the constant and replacing the PDFs and PMF in the above likelihood function and taking natural log thereafter we get,
\begin{equation*}
\begin{aligned}
    \mathcal{L}_1(\theta) = & \displaystyle -n_{1}~\mu + ln(\mu) \sum_{i \in S_1} (r_i-1)-\sum_{i \in S_1} ln((r_i-1)!) + n_1~ ln(p)+ \sum_{\substack{i\in S_{1}}}\biggl(\frac{r_{i}+1}{2}\biggr)~ ln(\lambda_1)\\
                    & -\lambda_1 \displaystyle\sum_{\substack{i\in S_{1}}}\sum_{j=1}^{\frac{r_{i}+1}{2}}x^1_{{i}j} + \sum_{\substack{i\in S_{1}}}\biggl(\frac{r_{i}-1}{2}\biggr)~ ln(\lambda_2) -\lambda_2 \sum_{\substack{i\in S_{1}}}\sum_{j=1}^{\frac{r_{i}-1}{2}}x^2_{{i}j}. 
\end{aligned}
\end{equation*}

\noindent \textbf{Situation 2}: The epoch starts with the event ``running without alert'' and the number of events $r_i$ is even.  Similar to  Situation~1, the log likelihood function here is 
\begin{equation*}\label{eq12}
\begin{aligned}
    \mathcal{L}_2(\theta) = & \displaystyle -n_{2}~\mu + ln(\mu) \sum_{i \in S_2} (r_i-1)-\sum_{i \in S_2} ln((r_i-1)!)+ n_2~ ln(p) +  \sum_{\substack{i\in S_{2}}}\biggl( \frac{r_{i}}{2}\biggr)~ ln(\lambda_1)\\
                    & -\lambda_1 \displaystyle\sum_{\substack{i\in S_{2}}}\sum_{j=1}^{\frac{r_{i}}{2}}x^1_{{ij}} + \sum_{\substack{i\in S_{2}}}^{n_2}\biggl(\frac{r_{i}}{2}\biggr)~ ln(\lambda_2) -\lambda_2 \sum_{\substack{i\in S_{2}}}\sum_{j=1}^{\frac{r_{i}}{2}}x^2_{ij}.
\end{aligned}
\end{equation*}

\noindent \textbf{Situation 3}: The epoch starts with the event ``running with alert'' and the number of events $r_i$ is odd. The log likelihood function in this situation is
\begin{equation*}\label{eq13}
\begin{aligned}
    \mathcal{L}_3(\theta) = & \displaystyle -n_{3}~\mu + ln(\mu) \sum_{i \in S_3} (r_i-1)-\sum_{i \in S_3} ln((r_i-1)!)+ n_3~ ln(1-p)+ \sum_{\substack{i\in S_{3}}}\biggl(\frac{r_{i}-1}{2}\biggr)~ ln(\lambda_1)\\
                    & -\lambda_1 \displaystyle\sum_{\substack{i\in S_{3}}}\sum_{j=1}^{\frac{r_{i}-1}{2}}x^1_{{i}j} + \sum_{\substack{i\in S_{3}}}\biggl(\frac{r_{i}+1}{2}\biggr)~ ln(\lambda_2) -\lambda_2 \sum_{\substack{i\in S_{3}}}\sum_{j=1}^{\frac{r_{i}+1}{2}}x^2_{{i}j}.
\end{aligned}
\end{equation*}

\noindent \textbf{Situation 4}: The epoch starts with the event ``running with alert'' and the number of events $r_i$ is even. Similarly, the log likelihood function here  is 
\begin{equation*}\label{eq14}
\begin{aligned}
    \mathcal{L}_4(\theta) = & \displaystyle -n_{4}~\mu + ln(\mu) \sum_{i \in S_4} (r_i-1)-\sum_{i \in S_4} ln((r_i-1)!)+ n_4 ~ ln(1-p) +  \sum_{\substack{i\in S_{4}}}\biggl(\frac{r_{i}}{2}\biggr)~ ln(\lambda_1)\\
                    & -\lambda_1 \displaystyle\sum_{\substack{i\in S_{4}}}\sum_{j=1}^{\frac{r_{i}}{2}}x^1_{{ij}} + \sum_{\substack{i\in S_{4}}}\biggl(\frac{r_{i}}{2}\biggr) ~ ln(\lambda_2) -\lambda_2 \sum_{\substack{i\in S_{4}}}\sum_{j=1}^{\frac{r_{i}}{2}}x^2_{{i}j}.
\end{aligned}
\end{equation*}


\section*{{Appendix A2: Derivation of MLEs of the base model}}
{From Equation (\ref{log_likelihood}), the log likelihood function can be expressed as
\begin{eqnarray*}
\mathcal{L}(\theta)&=& \mathcal{L}_1(\theta) +\mathcal{L}_2(\theta) +\mathcal{L}_3(\theta) +\mathcal{L}_4(\theta)\\
&= & -n \mu + ln (\mu) \sum_{l=1}^4 \sum_{i \in S_l} (r_{i}-1) + ln (\lambda_1) \sum_{l=1}^4\sum_{i \in S_l} \Big(\frac{r_i+a_l^1}{2}\Big)-\lambda_1 \sum_{l=1}^4 \sum_{i \in S_l} \sum_{j=1}^{\frac{r_i+a_l^1}{2}} x_{ij}^1\\
& &+ln (\lambda_2) \sum_{l=1}^4\sum_{i \in S_l} \Big(\frac{r_i+a_l^2}{2}\Big)-\lambda_2 \sum_{l=1}^4 \sum_{i \in S_l} \sum_{j=1}^{\frac{r_i+a_l^2}{2}} x_{ij}^2\\ 
&& +(n_1+n_2)~ ln(p) + (n_3+n_4)~ ln(1-p),
\end{eqnarray*}
where, for $s=1,2$, 
$$a_l^s=
\begin{cases}
(-1)^{s+1}, \text{~if $l=1$},\\
0, \text{~if $l=2,4$},\\
(-1)^s, \text{~if $l=3$}.\\
\end{cases}
$$
}
{The first order partial derivatives of $\mathcal{L}(\theta)$ with respective to $\lambda_1, \lambda_2,\mu$ and $p$ are separable,
\begin{eqnarray}
\frac{\partial \mathcal{L}(\theta)}{\partial \lambda_1}&=& \frac{\sum_{l=1}^4\sum_{i \in S_l} \Big(\frac{r_i+a_l^1}{2}\Big)}{\lambda_1}-\sum_{l=1}^4 \sum_{i \in S_l} \sum_{j=1}^{\frac{r_i+a_l^1}{2}} x_{ij}^1,\nonumber\\
\frac{\partial \mathcal{L}(\theta)}{\partial \lambda_2}&=&\frac{\sum_{l=1}^4\sum_{i \in S_l} \Big(\frac{r_i+a_l^2}{2}\Big)}{\lambda_2}-\sum_{l=1}^4 \sum_{i \in S_l} \sum_{j=1}^{\frac{r_i+a_l^2}{2}} x_{ij}^2,\nonumber\\
\frac{\partial \mathcal{L}(\theta)}{\partial \mu}&=&-n +\frac{\sum_{l=1}^4 \sum_{i \in S_l} (r_{i}-1)}{\mu},\nonumber\\
\frac{\partial \mathcal{L}(\theta)}{\partial p}&=&\frac{n_1+n_2}{p}-\frac{n_3+n_4}{1-p}.\nonumber
\end{eqnarray}
The normal equations are solved for $\lambda_1, \lambda_2, \mu, p$ giving explicit closed form expressions presented in Equation~(\ref{mles}). The second order partial derivatives of $\mathcal{L}(\theta)$ with respective to $\lambda_1, \lambda_2,\mu$ and $p$ evaluated at these solutions are negative, 
\begin{eqnarray}
\frac{\partial^2 \mathcal{L}(\theta)}{\partial {\lambda_1^2}}&=& -\frac{\sum_{l=1}^4\sum_{i \in S_l} \Big(\frac{r_i+a_l^1}{2}\Big)}{\lambda_1^2}<0,\nonumber\\
\frac{\partial^2 \mathcal{L}(\theta)}{\partial {\lambda_2^2}}&=& -\frac{\sum_{l=1}^4\sum_{i \in S_l} \Big(\frac{r_i+a_l^2}{2}\Big)}{\lambda_2^2}<0,\nonumber\\
\frac{\partial^2 \mathcal{L}(\theta)}{\partial {\mu^2}}&=& -\frac{\sum_{l=1}^4 \sum_{i \in S_l} (r_{i}-1)}{\mu^2}<0,\nonumber\\
\frac{\partial^2 \mathcal{L}(\theta)}{\partial {p^2}}&=&-\frac{n_1+n_2}{p^2}-\frac{n_3+n_4}{(1-p)^2}<0\nonumber.
\end{eqnarray}
Therefore, the exact solutions in Equation~(\ref{mles}) are the MLEs.}


\section*{{Appendix A3: Bootstrap confidence interval computation}}
{The following steps summarize the algorithm for computing parametric bootstrap confidence interval for ${\lambda}_{1}, {\lambda}_{2}, {p}$, and ${\mu}$.}
\begin{itemize}
\item {Step-1: Use original data to find the parameter estimates $\hat{\lambda}_{1}, \hat{\lambda}_{2}, \hat{p}$, and $\hat{\mu}$ (using the expressions in Equation~(\ref{mles})).}

\item {for ($m=1,2,\cdots,M$)}

\begin{itemize}
\item {Step-2: Use shifted-Poisson distribution with $\hat{\mu}$ to draw $r_1, r_2, \cdots, r_n$.}

\item {Step-3: Within each epoch, use Bernoulli distribution with $\hat{p}$ to decide whether the first event will be ``running without alert" or ``running with alert".}

\item {Step-4: Use exponential distributions with parameters $\hat{\lambda}_1$ and $\hat{\lambda}_2$ for generating the event duration for $r_i$ events in the $i$-th epoch.}

\item {Step-5: Use the expressions in Equation~(\ref{mles}) to find the parameters viz.  $\hat{\lambda}_{1m}, \hat{\lambda}_{2m}, \hat{p}_m$, and $\hat{\mu}_{m}$ from the dataset generated in Steps 2-4. }
\end{itemize}

\item {Step-6: Finally, $100(1-\alpha)\%$ bootstrap confidence interval for any $\eta \in \{\lambda_1, \lambda_2, p, \mu\}$ is given by 
$$\Big[\hat{\eta}_{([M\frac{\alpha}{2}])},\ \hat{\eta}_{([M(1-\frac{\alpha}{2})]}\Big],$$
where, $[x]$ represents the greatest integer function, and $\hat{\eta}_{(t)}$ denotes the $t$-th percentile of $\{\hat{\eta}_{m}, m = 1, 2, ..., M\}$.}
\end{itemize}


\section*{Appendix~A4: Regression model}

The GLM models presented in Equation~(\ref{eq:glm-params}) capture the dependence of 22 sensors on $\lambda_1, \lambda_{2}$ and $\mu$ via the regression coefficient vectors $\beta, \gamma$ and $\eta$, respectively. That is, for Situation~$l$, the log-likelihood can be written as
\begin{equation*} 
\mathcal{L}_l(\theta)=\mathcal{L}_l(\beta, \gamma, \eta, p)= g_{1l}(\beta) + g_{2l}(\gamma) + g_{3l}(\eta) +g_{4l}(p), 
\end{equation*}
where, $g_{1l}(\beta)$, $g_{2l}(\gamma)$, $g_{3l}(\eta)$ and $g_{4l}(p)$  are  functions of unknown quantities $\beta$, $\gamma$, $\eta$ and $p$. 
%
Thus, $g_{1l}(\beta)$ represents $\sum_{i\in S_{1}}\biggl(\frac{r_{i}+1}{2}\biggr)  ln(\lambda_{1i}) - \displaystyle\sum_{i\in S_{1}}\sum_{j=1}^{\frac{r_{i}+1}{2}}x^1_{{i}j} \lambda_{1i}$, and can be written as:
\begin{eqnarray*}
g_{1l}(\beta)&=&\sum_{i \in S_l} \Big(\frac{r_i+a_l^1}{2}\Big) \Big(\beta_0+\sum_{k=1}^m F_{ki} \beta_k\Big)-\sum_{i \in S_l} \sum_{j=1}^{\frac{r_i+a_l^1}{2}} x_{ij}^1~ exp\Big(\beta_0+\sum_{k=1}^m F_{ki} \beta_k\Big),
\end{eqnarray*}
$g_{2l}(\gamma)$ corresponds to $\sum_{i\in S_{1}}\biggl(\frac{r_{i}-1}{2}\biggr) ln(\lambda_{2i}) - \displaystyle\sum_{i\in S_{1}}\sum_{j=1}^{\frac{r_{i}-1}{2}}x^2_{{i}j} \lambda_{2i}$, and can be written as:
\begin{eqnarray*}
g_{2l}(\gamma)&=& \sum_{i \in S_l} \Big(\frac{r_i+a_l^2}{2}\Big) \Big(\gamma_0+\sum_{k=1}^m F_{ki} \gamma_k\Big)- \sum_{i \in S_l} \sum_{j=1}^{\frac{r_i+a_l^2}{2}} x_{ij}^2~ exp\Big(\gamma_0+\sum_{k=1}^m F_{ki} \gamma_k\Big),
\end{eqnarray*}
$g_{3l}(\eta)$ corresponds to  $-\sum_{i\in S_{1}}\mu_i + \displaystyle\sum_{i\in S_{1}}(r_i-1)ln(\mu_i)$ from $\mathcal{L}_l(\theta)$, and can be written as:
\begin{eqnarray*}
g_{3l}(\eta)&=& -\sum_{i \in S_l}exp\Big(\eta_0+\sum_{k=1}^m F_{ki} \eta_k\Big)+\sum_{i \in S_l} (r_{i}-1)\Big(\eta_0+\sum_{k=1}^m F_{ki} \eta_k\Big), 
\end{eqnarray*}
where,
$
a_l=\begin{cases}
1, \text{~if}~ l=1,\\
0, \text{~if}~ l=2,4,\\
-1, \text{~if}~ l=3.
\end{cases}
$

\section*{{Appendix A5: Derivation of MLEs of the regression model}}
{Using Equation~(\ref{Situation s log likelihood glm}) the log likelihood of the regression model can be expressed as}
{
\begin{eqnarray*}
\mathcal{L}(\eta, \beta, \gamma, p)&= & -\sum_{l=1}^4\sum_{i \in S_l}\Big[exp\Big(\eta_0+\sum_{k=1}^m F_{ki} \eta_k\Big)+ (r_{i}-1)\Big(\eta_0+\sum_{k=1}^m F_{ki} \eta_k\Big)\Big]\\
&&+\sum_{l=1}^4\sum_{i \in S_l} \Big[\Big(\frac{r_i+a_l^1}{2}\Big) \Big(\beta_0+\sum_{k=1}^m F_{ki} \beta_k\Big)-\sum_{j=1}^{\frac{r_i+a_l^1}{2}} x_{ij}^1~ exp\Big(\beta_0+\sum_{k=1}^m F_{ki} \beta_k\Big)\Big]\\
&&+\sum_{l=1}^4\sum_{i \in S_l} \Big[\Big(\frac{r_i+a_l^2}{2}\Big) \Big(\gamma_0+\sum_{k=1}^m F_{ki} \gamma_k\Big)-\sum_{j=1}^{\frac{r_i+a_l^2}{2}} x_{ij}^2~ exp\Big(\gamma_0+\sum_{k=1}^m F_{ki} \gamma_k\Big)\Big]\\
&&+(n_1+n_2)~ ln(p) + (n_3+n_4)~   ln(1-p).
\end{eqnarray*}}
{The first order partial derivative of $\mathcal{L}(\eta, \beta, \gamma, p)$ with respect to the $j$-th ($j=0,1,\dots m$) component of $\eta, \beta, \gamma$ are as follows.}
{\begin{eqnarray*}
\frac{\partial \mathcal{L}(\eta, \beta, \gamma, p)}{\partial {\eta_j}}=\begin{cases}
-\sum_{l=1}^4\sum_{i \in S_l}\Big[exp\Big(\eta_0+\sum_{k=1}^m F_{ki} \eta_k\Big)+ (r_{i}-1)  \Big], \text{~if $j=0$,}\\
-\sum_{l=1}^4\sum_{i \in S_l}\Big[exp\Big(\eta_0+\sum_{k=1}^m F_{ki} \eta_k\Big)F_{ji}+ (r_{i}-1) F_{ji} \Big], \text{~if $j=1,2,\cdots, m$.}
\end{cases}\\
\frac{\partial \mathcal{L}(\eta, \beta, \gamma, p)}{\partial {\beta_j}}=\begin{cases}
\sum_{l=1}^4\sum_{i \in S_l} \Big[\Big(\frac{r_i+a_l^1}{2}\Big)  -\sum_{j=1}^{\frac{r_i+a_l^1}{2}} x_{ij}^1~ exp\Big(\beta_0+\sum_{k=1}^m F_{ki} \beta_k\Big)\Big], \text{~if $j=0$,}\\
\sum_{l=1}^4\sum_{i \in S_l} \Big[\Big(\frac{r_i+a_l^1}{2}\Big)  F_{ji} -\sum_{j=1}^{\frac{r_i+a_l^1}{2}} x_{ij}^1~ exp\Big(\beta_0+\sum_{k=1}^m F_{ki} \beta_k\Big)F_{ji}\Big], \text{~if $j=1,2,\cdots, m$.}
\end{cases}\\
\frac{\partial \mathcal{L}(\eta, \beta, \gamma, p)}{\partial {\gamma_j}}=\begin{cases}
\sum_{l=1}^4\sum_{i \in S_l} \Big[\Big(\frac{r_i+a_l^2}{2}\Big)   -\sum_{j=1}^{\frac{r_i+a_l^2}{2}} x_{ij}^2~ exp\Big(\gamma_0+\sum_{k=1}^m F_{ki} \gamma_k\Big)\Big], \text{~if $j=0$,}\\
\sum_{l=1}^4\sum_{i \in S_l} \Big[\Big(\frac{r_i+a_l^2}{2}\Big)  F_{ji} -\sum_{j=1}^{\frac{r_i+a_l^2}{2}} x_{ij}^2~ exp\Big(\gamma_0+\sum_{k=1}^m F_{ki} \gamma_k\Big)F_{ji}\Big], \text{~if $j=1,2,\cdots, m$.}
\end{cases}
\end{eqnarray*}}
{To obtain MLEs of the parameters, the normal equations are solved for $\eta_j, \beta_j, \gamma_j$. Clearly, the second order partial derivatives of $\mathcal{L}(\eta, \beta, \gamma, p)$ with respect to $\eta_j, \beta_j, \gamma_j$ evaluated at these solutions are negative.}
{
\begin{eqnarray*}
\frac{\partial^2 \mathcal{L}(\eta, \beta, \gamma, p)}{\partial {\eta_j^2}}=\begin{cases}
- \sum_{l=1}^4\sum_{i \in S_l} exp\Big(\eta_0+\sum_{k=1}^m F_{ki} \eta_k\Big), \text{~ if $j=0$,}\\
- \sum_{l=1}^4\sum_{i \in S_l} exp\Big(\eta_0+\sum_{k=1}^m F_{ki} \eta_k\Big)F_{ji}^2, \text{~ if $j=1,2,\cdots, m$.}
\end{cases}\\
\frac{\partial^2 \mathcal{L}(\eta, \beta, \gamma, p)}{\partial {\beta_j^2}}=\begin{cases}
- \sum_{l=1}^4\sum_{i \in S_l} \sum_{j=1}^{\frac{r_i+a_l^1}{2}} x_{ij}^1~ exp\Big(\beta_0+\sum_{k=1}^m F_{ki} \beta_k\Big), \text{~ if $j=0$,}\\
- \sum_{l=1}^4\sum_{i \in S_l} \sum_{j=1}^{\frac{r_i+a_l^1}{2}} x_{ij}^1~ exp\Big(\beta_0+\sum_{k=1}^m F_{ki} \beta_k\Big)F_{ji}^2, \text{~ if $j=1,2,\cdots, m$.}
\end{cases}\\
\frac{\partial^2 \mathcal{L}(\eta, \beta, \gamma, p)}{\partial {\gamma_j^2}}=\begin{cases}
- \sum_{l=1}^4\sum_{i \in S_l} \sum_{j=1}^{\frac{r_i+a_l^2}{2}} x_{ij}^2~ exp\Big(\gamma_0+\sum_{k=1}^m F_{ki} \gamma_k\Big), \text{~ if $j=0$,}\\
- \sum_{l=1}^4\sum_{i \in S_l} \sum_{j=1}^{\frac{r_i+a_l^2}{2}} x_{ij}^2~ exp\Big(\gamma_0+\sum_{k=1}^m F_{ki} \gamma_k\Big)F_{ji}^2, \text{~ if $j=1,2,\cdots, m$.}
\end{cases}
\end{eqnarray*}}
{ Hence, the roots are MLEs, but since these equations are non-linear in nature which cannot be solved analytically, we used ``optim" function in R software for finding these MLEs.}

\section*{Appendix~A6: Important sensors}

When random forest (RF) models were fitted to M1-M5 in Section~5, the top ten important predictors (IM sensors) are not necessarily common.  Model wise list is presented in Table~\ref{classificatiom model cv}. Table~\ref{description of selected cv} describes the union of these top ten important predictors.

\begin{table}[!h]\centering
	\caption[Description of selected sensors]{Description of selected sensors}
	\label{description of selected cv}
	\begin{tabular}{|m{1in}|m{3.5in}|}
		
		\hline \multicolumn{1}{|c|}{\textbf{sensors}} & \multicolumn{1}{c|}{\textbf{Description}} \\ \hline

		$F_{04}$, $F_{05}$, $F_{07}$, $F_{15}$, $F_{16}$ & Mold Surface Temp Cavity: The temperature of the mold's surface in the $1^{st}$, $10^{th}$, $12^{th}$, $5^{th}$ and $6^{th}$ cavities respectively, of the IM machine.\\ \hline
		
		$F_{21}$, $F_{22}$, $F_{24}$, $F_{27}$, $F_{29}$, $F_{30}$, $F_{31}$, $F_{32}$, $F_{34}$, $F_{35}$  &  Cooling Rate PostGate Pressure Drop Cavity: The cooling rate at which the pressure decreases per second in the $10^{th}$, $11^{th}$, $13^{th}$, $2^{nd}$, $4^{th}$, $5^{th}$, $6^{th}$, $7^{th}$, $9^{th}$ and $15^{th}$ respectively, of the mold after the gate, as the plastic material in that cavity cools and solidifies. Monitoring this parameter is important for ensuring proper cooling and solidification of the plastic material to achieve the desired part quality and dimensional stability. \\ \hline

		$F_{44}$, $F_{51}$ & Peak PostGate Cavity: Peak pressure at post gate rear cavity $3^{rd}$ and $15^{th}$ respectively. \\ \hline

		$F_{58}$,~$F_{66}$, $F_{67}$ & Injection Integral PostGate Cavity: Calculated value or measurement that quantifies the behavior in the $16^{th}$, $9^{th}$ and $15^{th}$ cavity respectively of the mold post rear gate during the IM process. \\  \hline  
		
		$F_{68}$          & Injection Integral of Molding.\\ \hline
		
		$F_{69}$          & The Injection Fill Time (Sec).\\ \hline
	\end{tabular}
\end{table}

\newpage
\section*{Appendix~A7: Prediction of time to fail}
 The expected time to fail can be expressed as
\begin{eqnarray*}
E[\text{Time to fail}]&=&E_R\left(E[\text{Time to fail} | R=r]\right) 
=\sum_{r=1}^{\infty} E[\text{Time to fail} |R=r] P(R=r) \nonumber\\
&=& \sum_{r\in \{1,3,5,\cdots\}} E[\text{Time to fail} |R=r] P(R=r) \nonumber\\
&&+  \sum_{r\in \{2,4,6,\cdots\}} E[\text{Time to fail} | R=r] P(R=r) 
\end{eqnarray*}
where $R$ is the number of events in an epoch. Now, for $r \in \{1,3,5,\cdots,\}$, 
\begin{eqnarray}
E[\text{Time to fail | R=r}]&=& E[\text{Time to fail}~ |~ R=r, Type=1]~ P(Type=1 | R=r )\nonumber\\
& &+ E[\text{Time to fail}~ |~ R=r, Type=2]~ P(Type=2 | R=r )\nonumber\\
&=& \bigg[ \frac{r+1}{2\lambda_1} +  \frac{r-1}{2\lambda_2}\bigg] p + \bigg[ \frac{r-1}{2\lambda_1} + \frac{r+1}{2\lambda_2}\bigg] (1-p),\nonumber
\label{sum_1}
\end{eqnarray}
where $Type=1$ indicates that the epoch starts with the event ``running without alert" and $Type=2$ indicates that the epoch starts with the event ``running with alert". 

Hence, 
\begin{eqnarray}\label{odd_sum}
&&\sum\limits_{r\in \{1,3,5,\cdots\}} E[\text{Time to fail}~ |~R=r]~ P(R=r)\nonumber\\
&&=  \frac{p}{2\lambda_1} \sum\limits_{r\in \{1,3,5,\cdots\}} (r+1) e^{-\mu} \frac{\mu^{r-1}}{(r-1)!} 
+ \frac{p}{2\lambda_2} \sum\limits_{r\in \{1,3,5,\cdots\}} (r-1) e^{-\mu} \frac{\mu^{r-1}}{(r-1)!} \nonumber\\
& &+\frac{1-p}{2\lambda_1} \sum\limits_{r\in \{1,3,5,\cdots\}} (r-1) e^{-\mu} \frac{\mu^{r-1}}{(r-1)!}
+ \frac{1-p}{2\lambda_2} \sum\limits_{r\in \{1,3,5,\cdots\}} (r+1) e^{-\mu} \frac{\mu^{r-1}}{(r-1)!}\nonumber\\
&&=\frac{\mu}{4} (1-e^{-2\mu}) \left[\frac{1}{\lambda_1}+\frac{1}{\lambda_2}\right]
+\frac{1}{2} (1+e^{-2\mu}) \left[\frac{p}{\lambda_1}+\frac{1-p}{\lambda_2}\right]. 
\end{eqnarray}

For $r \in \{2,4,6,\cdots,\}$, 
\begin{eqnarray}
E[\text{Time to fail}~ |~ R=r]&=& E[\text{Time to fail}~ |~ R=r, Type=1]~ P(Type=1 | R=r )\nonumber\\
& &+ E[\text{Time to fail}~ |~ R=r, Type=2]~ P(Type=2 | R=r )\nonumber\\
&=& \bigg[ \frac{r}{2\lambda_1} +  \frac{r}{2\lambda_2}\bigg] p + \bigg[ \frac{r}{2\lambda_1} +  \frac{r}{2\lambda_2}\bigg] (1-p) = \frac{r}{2} \bigg[\frac{1}{\lambda_1}  + \frac{1}{\lambda_2} \bigg].\nonumber
\end{eqnarray}

Hence, \begin{eqnarray}\label{even_sum}
& &\sum\limits_{r\in \{2,4,6,\cdots\}} E[\text{Time to fail}~ |~R=r]~ P(R=r)\nonumber\\
&&=\frac{1}{2}\left(\frac{1}{\lambda_1}+\frac{1}{\lambda_2}\right) \sum\limits_{r \in \{2,4,6,\cdots\}} r e^{-\mu} \frac{\mu^{r-1}}{(r-1)!}\nonumber\\
&&= \frac{1}{4}\left(\frac{1}{\lambda_1}+\frac{1}{\lambda_2}\right)\big[\mu(1+e^{-2\mu})
+(1-e^{-2\mu})\big].
\end{eqnarray}
Thus, the expected time to fail of an epoch is obtained by adding Equations~(\ref{odd_sum}) and (\ref{even_sum}).

\end{document}